\documentclass[runningheads]{llncs}
\usepackage[T1]{fontenc}
\usepackage[table]{xcolor}
\usepackage[hidelinks]{hyperref}
\usepackage{orcidlink}
\usepackage{xparse}
\usepackage{graphicx}
\usepackage{todonotes}
\usepackage{adjustbox}
\usepackage{amsmath}
\usepackage{amssymb}
\usepackage{enumitem}
\usepackage{wrapfig}
\usepackage{caption}
\usepackage{subcaption}
\usepackage{pslatex}
\usepackage{tablefootnote}
\usepackage{multirow}
\usepackage[ruled,linesnumbered,noend]{algorithm2e}
\usepackage{stmaryrd}
\usepackage{mathtools}
\usepackage{tikz}
\usepackage{cleveref}
\usepackage{changepage}
\usepackage{nicefrac}
\usepackage{marvosym}

\usetikzlibrary{shapes,shapes.geometric,arrows,fit,calc,automata,positioning}
\tikzset{elliptic state/.style={draw, ellipse, inner sep=1mm}}
\tikzset{node distance=2.5cm}
\tikzset{every state/.style={minimum size=0.7cm}}
\tikzset{rectangular state/.style={draw, rounded rectangle, inner sep=2mm, minimum height=0.7cm}}
\usetikzlibrary {decorations.pathreplacing}

\newcommand{\algout}[1]{#1'}
\newcommand{\crightarrow}[1]{\hspace{0.25ex}\raisebox{0.1ex}{\scriptsize$\relbar$\hspace*{-0.5ex}#1\hspace*{-0.5ex}$\rightarrow$}\hspace{0.25ex}}
\newcommand{\transition}[3]{#1 \crightarrow{\textnormal{\{}$#2$\textnormal{\}}} #3}

\makeatletter
\newenvironment{algorithmNoBottom}
{
  \renewcommand*{\@algocf@post@ruled}{}
  \begin{algorithm}%
}
{
  \end{algorithm}%
  \RestyleAlgo{ruled}
}

\newenvironment{algorithmNoTop}
{%
  \RestyleAlgo{plain}
  \begin{algorithm}%
}
{%
  \end{algorithm}%
  \vspace*{-1em}
  \hrule height 1pt
  \RestyleAlgo{ruled}
  \vspace*{1em}
}
\let\oldnl\nl
\newcommand{\nonl}{\renewcommand{\nl}{\let\nl\oldnl}}
\makeatother

\begin{document}
\title{Automata Size Reduction by Procedure Finding
\thanks{
This work was supported by the Czech Science Foundation project GA23-07565S; the Czech Ministry of Education, Youth and Sports ERC.CZ project LL1908; and
the FIT BUT internal project FIT-S-23-8151.
}
}

\author{Michal Šedý \and Lukáš Holík}
\author{Michal Šedý\inst{1, 2}\textsuperscript{(\Letter)}\orcidlink{0009-0004-0091-0546} \and Lukáš Holík\inst{1, 2}\orcidlink{0000-0001-6957-1651}}
\authorrunning{M. Šedý, L. Holík}
\institute{Faculty of Information Technology, Brno University of Technology, \\Brno, Czech Republic\\
\and The Technical Faculty of IT and Design, Aalborg University, \\Aalborg {\O}st, Denmark\\
\email{\{medy,lukasholik\}@cs.aau.dk}}
\maketitle

\begin{abstract}
We introduce a novel paradigm for reducing the size of finite automata by compressing repeating sub-graphs. These repeating sub-graphs can be viewed as invocations of a single procedure.
Instead of representing each invocation explicitly, they can be replaced by a single procedure that uses a small runtime memory to remember the call context.
We elaborate on the technical details of a~basic implementation of this idea, where the memory used by the procedures is a~simple finite-state register. We propose methods for identifying repetitive sub-graphs, collapsing them into procedures, and measuring the resulting reduction in automata size. Already this basic implementation of reduction by procedure finding yields practically relevant results, particularly in the context of FPGA-accelerated pattern matching, where automata size is a primary bottleneck. We achieve a size reduction of up to 70\% in automata that had already been minimized using existing advanced methods.
\end{abstract}

\section{Introduction}
One of the primary bottlenecks in applications of finite automata is their size.
Especially when they are manipulated by boolean operations or used to represent syntactic features such as counting, the problem of state space explosion becomes the main bottleneck.
Research on succinctly representing automata and reducing their size has a~long history.
Succinct representations of automata that use syntactic features such as non-determinism, alternation, or various kinds of memories (counters, registers, stacks) are used.
Size reduction methods have been extensively studied, based mainly on state merging and transition pruning \cite{Oldest_Merge,Simulation_based_minimization,Lorenzo_prunning_saturation,On_nfa_reduction,NFA_min}, as well as SAT solving or other techniques \cite{SAT_omega,ApproxRed,SAT_NFA,SAT_molnarova,SAT_sedy}.

The reduction methods normally stay within the original syntactic class of automata,
and the succinct syntactic features can only be used if present explicitly in the problem formulation, e.g., an automaton with counter is parsed from a regular expression with counters, an alternating automaton is generated from boolean operations over automata, and a push-down automaton is compiled from a system with procedures.

In this paper, we infer succinct syntactic features in nondeterministic finite automata (NFAs), namely
procedures, that were \emph{not} explicitly present. We identify repeating sub-graphs and replace them by procedure calls.
The automaton size decreases this way since the procedure is represented only once.
Procedures are implemented via an additional memory which they use to remember the call context.

In this first take on the idea, we choose a single finite domain register as the simplest possible implementation of the memory.
A procedure invocation is started by a~call transition that stores a value identifying the call context in a register. The value is checked on the return from the procedure and on transitions that should be enabled only in some contexts.
We favor the simplicity of the finite domain register over some perhaps more natural choices of implementing the procedure context, such as push-down or hierarchical automata.
This simplicity should make it relatively easy to extend existing NFA based methods and tools to automata with a register.
We particularly experimented with automata used in hardware accelerated regular pattern matching \cite{FPGA_based_network_scaning} where the size of the automata is a primary concern.
We believe that the basic automata algorithms, such as boolean operations and techniques used in areas such as regular model checking or string constraint solving, and their implementations in libraries such as Mata or Brics \cite{mata,brics},
should be extensible to work with this implementation of procedures as well.

\begin{wrapfigure}[23]{r}{0.33\linewidth}
    \vspace*{-2.25em}
    \begin{minipage}{\textwidth}
        \begin{subfigure}{0.3\textwidth}
            \resizebox{0.9\textwidth}{!}{
                \centering
                \begin{tikzpicture}[node distance=1.75cm]
                    \node[state, initial] (q0) {$q_0$};
                    \node[state, above=0.5cm of q0] (q1) {$q_1$};
                    \node[state, below=0.5cm of q0] (q2) {$q_2$};
                    \node[state, right of=q1] (q3) {$q_3$};
                    \node[state, right of=q2] (q4) {$q_4$};
                    \node[state, right of=q3] (q5) {$q_5$};
                    \node[state, right of=q4] (q6) {$q_6$};
                    \node[state, accepting, below=0.5cm of q5] (q7) {$q_7$};

                    \path[->]
                            (q0) edge[left, pos=0.45] node{$x$} (q1)
                            (q0) edge[left, pos=0.45] node{$y$} (q2)
                            (q1) edge[above] node{$a$} (q3)
                            (q2) edge[below] node{$a$} (q4)
                            (q3) edge[above] node{$a$} (q5)
                            (q3) edge[loop below] node{$c$} (q3)
                            (q4) edge[below] node{$a,b$} (q6)
                            (q5) edge[right, pos=0.45] node{$x$} (q7)
                            (q6) edge[right, pos=0.45] node{$y$} (q7);
                \end{tikzpicture}
                }
            \vspace*{-0.4em}
            \caption{Before the reduction.}
        \end{subfigure}

        \vspace*{0.6em}

        \begin{subfigure}{0.31\textwidth}
            \centering
            \hspace*{1.5ex}
            \begin{adjustbox}{max width=0.19\textwidth}
                \begin{tikzpicture}[node distance=1.2cm]
                    \node[rectangular state] (q12) {$(q_1, q_2)$};
                    \node[rectangular state, below of=q12] (q34) {$(q_3, q_4)$};
                    \node[rectangular state, below of=q34] (q56) {$(q_5, q_6)$};

                    \path[->]
                            (q12) edge[right] node{$a$} (q34)
                            (q34) edge[right] node{$a$} (q56);
                \end{tikzpicture}
            \end{adjustbox}
            \vspace*{-0.2em}
            \caption{A part of the self-product identifying the similar sub-graphs (the infix \texttt{aa}).}
        \end{subfigure}

        \vspace*{0.5em}

        \begin{subfigure}{0.3\textwidth}
            \resizebox{\textwidth}{!}{
                \centering
                \begin{tikzpicture}[node distance=1.75cm]
                    \node[state, initial] (q0) {$q_0$};
                    \node[rectangular state, above=0.75cm of q0] (q12) {$q_{1,2}$};
                    \node[rectangular state, right of=q12] (q34) {$q_{3,4}$};
                    \node[rectangular state, right of=q34] (q56) {$q_{5,6}$};
                    \node[state, accepting, below=0.75cm of q56] (q7) {$q_7$};

                    \path[->]
                            (q0) edge[left, pos=0.45] node{$\begin{aligned}
                                                    x,\bot&/1\\[-1.5mm]
                                                    y,\bot&/2
                                                \end{aligned}$} (q12)
                            (q12) edge[above] node{$a,*/*$} (q34)
                            (q34) edge[above] node{$\begin{aligned}
                                                        a,*&/*\\[-1.5mm]
                                                        b,2&/2
                                                    \end{aligned}$} (q56)
                            (q34) edge[loop below] node{$c,1/1$} (q34)
                            (q56) edge[right, pos=0.45] node{$\begin{aligned}
                                                    x,1&/\bot\\[-1.5mm]
                                                    y,2&/\bot
                                                \end{aligned}$} (q7);
                \end{tikzpicture}
            }
            \caption{After the reduction.}
            \vspace{-0.4em}
        \end{subfigure}
    \end{minipage}
    \caption{Reduction of an\hspace{1.5ex}\break automaton compiled from\hspace{1.5ex}\break \texttt{(xac$^*$ax)+(ya\!(a+b)\!y)}.}
    \label{fig:aa}
    \vspace*{-3em}
\end{wrapfigure}

In order to reduce the size of an automaton, we first search for pairs of similar sub-graphs of the automaton. Technically, we explore the product of the automaton with itself
and we search in it for parts that represent overlayed sub-graphs of the automaton that are similar.
The search is essentially a greedy algorithm driven by an estimation of the size reduction that would be achieved by representing the overlayed automaton sub-graphs by a procedure.
To keep the search practically feasible, we only consider procedures with linear spanning trees. Repeating automata sub-graphs of this shape occur frequently in our benchmark.
For instance, they appear in automata representing regular expressions such as
{\texttt{(.*new XMLHttpRequest.*file://)|(.*file://.*new XML\break HttpRequest)}} from \cite{Snort_git}
that consists of two parts that may swap positions,
or in automata representing words that share a common infix, such as {\texttt{(xac$^*$ax)+(ya\!(a+b)\!y)}} shown in Figure \ref{fig:aa}.
The pair of two similar sub-graphs, seen as two invocations of the same procedure, is then replaced by the single procedure that can be called from the two contexts and remembers the context in the register, as shown on transitions $\transition{q_0}{x,\bot/1}{q_{1,2}}$ and $\transition{q_0}{y,\bot/2}{q_{1,2}}$ entering the procedure in Figure \ref{fig:aa}c.
It is important for the practical efficiency of the reduction that the two original procedure invocations do not need to be completely isomorphic. Transitions unique for only one of the invocations are permitted.
The procedure can use the value stored in the register to decide whether the transition can be used (as is the case of the transition $\transition{q_{3,4}}{b,2/2}{q_{5,6}}$ and the self-loop $\transition{q_{3,4}}{c,1/1}{q_{3,4}}$ in Figure \ref{fig:aa}c).

We experimented with automata generated from network filtering rules used in the network intrusion detection system Snort \cite{Snort}.
To optimize deployment in FPGA accelerated pattern matching \cite{FPGA_based_network_scaning}, where the size of the automata is a crucial bottleneck due to the limited space on the FPGA chip, these automata were reduced by advanced simulation-based reductions, implemented in the RABIT/Reduce tool \cite{RABIT}.
Yet, by additionally applying our approach, we achieved a further reduction of up-to 56\% in the number of states and up-to 60\% in the number of transitions, which can significantly improve the capabilities of the intrusion detection method (provided that an adequate encoding of the introduced automaton in the FPGA is found).
We then achieved similar results
on automata from other diverse sources,
namely, abstract regular model checking \cite{ARMC}
string constraint solver Z3-Noodler \cite{z3-noodler-git},
regular expressions for email validation from the RegexLib library \cite{RegexLib} and parametric regular expressions \cite{TACAS13}. Here we were able to reduce the number of states and transitions by up to 69\% and 74\%, respectively.

\subsubsection{Related Work.}
Techniques used to reduce NFAs have been studied for a long time. This includes mainly state merging \cite{Oldest_Merge,Simulation_based_minimization,On_nfa_reduction,NFA_min},
transition pruning \cite{Simulation_based_minimization,Lorenzo_prunning_saturation}, and also transition saturation \cite{Lorenzo_prunning_saturation}, based on computing the inclusion of languages on automata states and its approximations, such as simulation and bisimulation relations. There were also attempts to use SAT-solvers \cite{SAT_omega,ApproxRed,SAT_NFA,SAT_molnarova,SAT_sedy} and methods for finding truly minimal NFAs \cite{NFA_min}.
To the best of our knowledge, the idea of automatically inferring syntactic features such as registers for the sake of succinctness has been explored only minimally. We are only aware of works \cite{CFLOBDDs} and \cite{XFA}.
In \cite{XFA}, finite automata with scratch memory, similar to our register, are used in signature matching in network intrusion detection. The approach, although demonstrating that the idea can be practical, is however ad-hoc, hard wired into the network intrusion detection pipeline, works for only few simple forms of regular expressions, and requires parts of completely isomorphic regular expressions to be manually annotated.
The work \cite{CFLOBDDs} reduces the size of BDDs by collapsing equivalent sub-BDDs into a single procedure, with the call context used to determine the appropriate return point.
Although some of the ideas used there are similar to ours, the technique is quite far from our work, being strongly tied to the specifics of the BDD data structure, their determinism, binary alphabet, levels corresponding to variables, and acyclicity. It targets canonical representation of boolean functions that can be achieved by a deterministic inductive construction and requires complete isomorphism of sub-graphs represented as a procedure.

\section{Preliminaries}
We start by preliminaries on finite automata, after which we specify their extension with a single finite domain register we use for representing procedures.

\paragraph{Finite Automata.}
A \textit{nondeterministic finite automaton (NFA)} is a 5-tuple $A = (Q, \Sigma, \delta,\break I, F)$, where $Q$ is a~finite nonempty set of \emph{states}, $\Sigma$ is an \emph{input alphabet}, $\delta \subseteq Q \times \Sigma \times Q$ is a set of \emph{transitions} in the form $r \crightarrow{$\{a\}$} s$, where $r \in Q$ is a \emph{source state}, $a \in \Sigma$ is an \emph{input symbol}, and $s \in Q$ is a \emph{target state}, $I \subseteq Q$ is a~nonempty set of \emph{initial states}, and $F \subseteq Q$ is a set of \emph{final states}.
A \textit{run} over a word $w = a_1\cdot a_2\cdots a_n \in \Sigma^*$ is a sequence of transitions $\transition{r_0}{a_1}{r_1}, \transition{r_1}{a_2}{r_2}, \cdots, \transition{r_{n-1}}{a_n}{r_n}$, where $\transition{r_i}{a_{i+1}}{r_{i+1}} \in \delta$ for all $0 \leq i < n$. The run is \textit{accepting} if $r_0 \in I$ and $r_n \in F$. The \textit{language} $L(A) \subseteq \Sigma^*$ of $A$ is a set of words for which $A$ has an accepting run.

A \emph{sub-graph} of $A$ induced by a set of states $P\subseteq Q$ is pair $(P,\delta')$ where $\delta'\subseteq \delta$ contains the transitions of $\delta$ with both end-points in $P$.

\paragraph{Single-Finite-Register Automata.}
A \textit{single-finite-register automaton (SRA)} is a~6-tuple $A = (Q, \Sigma, \Gamma, \delta, I, F)$ such that $(Q, \Sigma \times \Gamma_{\hspace{-0.5ex}\bot, *} \times \Gamma_{\hspace{-0.5ex}\bot,*}, \delta, I, F)$ is an NFA, $\Gamma$ is a finite \emph{register alphabet} and $\Gamma_{\hspace{-0.5ex}\bot,*} = \Gamma \cup \{\bot, *\}$, with $\bot \notin \Gamma$ being a \textit{default register value} and $* \notin \Gamma$ being a \textit{wildcard}.
We denote transitions of the SRA as $r\crightarrow{$\{a,\alpha/\beta\}$}s$, where $a \in \Sigma$ is an \emph{input symbol}, $\alpha \in \Gamma_{\hspace{-0.5ex}\bot,*}$ is a \emph{test symbol}, and $\beta \in \Gamma_{\hspace{-0.5ex}\bot,*}$ a \emph{set symbol}.
A~\textit{configuration} of the SRA is a~tuple $(q, \alpha) \in Q \times \Gamma_{\hspace{-0.5ex}\bot}$.
The \textit{induced} NFA of $A$ is a 6-tuple $A' = (Q', \Sigma, \delta', I', F')$, where $Q' \subseteq Q \times \Gamma_{\hspace{-0.5ex}\bot}$ are configurations of SRA, $I' \subseteq I \times \{\bot\}$ are initial configurations, $F' \subseteq Q'$ are final configurations, and $\delta' = \{\transition{(r, \alpha)}{a}{(s,\beta)}\,|\, (\transition{r}{a,\alpha,\beta}{s} \in \delta) \lor (\alpha = \beta \land \transition{r}{a,*/*}{s} \in \delta )\}$ represents transitions between SRA's configurations.
Intuitively, a~transition $r \crightarrow{$\{a,\alpha/\beta\}$} s$ can be taken if the register value matches the test symbol $\alpha$, in which case the new register value is set to $\beta$. Furthermore, a transition $r \crightarrow{$\{a,*/*\}$} s$ can be taken regardless of the register value, and the register value remains unchanged.
We denote by $\mathit{Reach}(A)$ a set of configurations that can be reached by a run starting from $I'$.
The language of SRA $A$ can be defined as the language of the induced NFA, $L(A) = L(A')$.

\section{Procedures}
Intuitively, similar disjoint sub-graphs of an automaton can be seen as \emph{invocations} of the same \emph{procedure}. The two invocations can be replaced by the procedure, represented only once and accessible in the two original contexts.
When called, the procedure will remember the current context by storing a specific \textit{invocation Id symbol} in the SRA register.
The invocation symbol will be later used to determine where to return after the exit from the procedure and what invocation context specific transitions should be enabled.
The SRA with the newly introduced procedure call will be smaller than the original NFA, but the induced NFA will be isomorphic to the original.

For example, in Figure \ref{fig:aa}, the two similar sub-graphs (invocations) span over states $q_1, q_3, q_5$ and $q_2, q_4, q_6$. To distinguish between them, the invocation Ids $1$ and $2$ will be used. These symbols will be set on call transitions $\transition{q_0}{x,\bot/1}{q_{1,2}}$ and $\transition{q_0}{y,\bot/2}{q_{1,2}}$ and later used on internal transitions $\transition{q_{3,4}}{b,2/2}{q_{5,6}}$ and $\transition{q_{3,4}}{c,1/1}{q_{3,4}}$ and return transitions $\transition{q_{5,6}}{x,1/\bot}{q_{3,4}}$ and $\transition{q_{5,6}}{y,2/\bot}{q_{3,4}}$.

\paragraph{Procedures Formally.}
A procedure is formally defined as a sub-graph of a SRA $A = (Q, \Sigma, \Gamma, \delta, I, F)$.
It is constructed to represent multiple invocations in an original SRA\footnote{For each NFA, there is a language equivalent isomorphic SRA that does not utilize the register.}.
Since the procedure represents invocations (each distinguished by unique invocation Id),
it is important to track which Ids can appear in the register when the computation is in a particular state of the procedure. This leads to the concept of a \emph{signature of a~state}. Formally defined as $\mathit{Sig}(q) = \{\beta \in \Gamma \,|\, \exists q \in Q: (q, \beta) \in \mathit{Reach}(A)\}$.
For example, in Figure \ref{fig:aa}, the state $q_{1,2}$ with $\mathit{Sig}(q_{1,2}) = \{1, 2\}$ originates from the two states $q_1$ and $q_2$ belonging to invocations with Ids $1$ and $2$.

Since a procedure can be nested,
different states within a procedure may have different state signatures.
However, due to the way nested procedures are constructed, from the outermost to the innermost, we know that there must have been invocations that have formed the outermost procedure.
This implies the existence of a set $I \subseteq \Gamma$ of invocation Ids, corresponding to those invocations, that must appear in the signature of every state within the procedure. We refer to this set $I$ as the \emph{procedure signature}.

Now, we can define a procedure as a maximal set of states that share the same procedure signature $I$. Formally, a nonempty set $P \subseteq Q$ is the set of states of a procedure if and only if there exists a set $I \subseteq \Gamma$ of Ids such that $\forall q\in Q:I \subseteq \mathit{Sig}(q) \iff q \in P$.

The \emph{call transitions} of the procedure are the transitions from $Q\setminus P$ to $P$, the \emph{return transitions} lead from $P$ to $Q\setminus P$, and the \emph{internal transitions} lead from $P$ to $P$ (they are the transitions of the sub-graph, unlike call and return transitions).
In our construction, the call transitions will always set the register symbol, some internal transitions may test it, and the return transition always test the register and sets it to the default value $\bot$.
Our algorithm will only produce SRAs where the procedures are \emph{well nested}, meaning that the sets of internal transitions of two procedures are either disjoint or one is included in the other.

\section{Constructing a Procedure from Similarity Graph}

The automata reduction by procedure finding iterates two steps.
First, it identifies pairs of similar sub-graphs that are to be understood as procedure invocations (e.g., sub-graphs induced by states $q_1, q_3, q_5$ and $q_2, q_4, q_6$ in Figure \ref{fig:aa}a).
The search returns a~\emph{similarity graph} that defines a partial isomorphism between the two invocations (e.g., Figure \ref{fig:aa}b).
The two invocations are then replaced by a~single procedure, constructed from the similarity graph (e.g., Figure \ref{fig:aa}c).
We will first concentrate on the creation of a~procedure from a~given similarity graph.

Formally, a \textit{similarity graph} of SRA $A = (Q, \Sigma, \Gamma, \delta, I, F)$ relates two similar disjoint sub-graphs of $A$, that can be then represented as one procedure.
It is a directed graph $G = (V, E)$.
Its vertices $V \subseteq Q \setminus (I \cup F) \times Q\setminus(I \cup F)$ represent pairs of \textit{similar states}.%
\footnote{For simplicity, we exclude initial and final states from participating in the procedure based reduction. This is without loss of generality since every NFA can be easily transformed to an equivalent one with at most two states in $I\cup F$ (add a new initial state $i$, a final state $f$, and make $i$ also final if $\epsilon\in L(A)$, and every transition to $f$/from $i$ for every original transition to/from a final/an initial state, respectively). The exclusion of the two states is negligible.}
Let $\pi_k$ be a~projection onto the $k$-th element in a set of pairs.
Then $\pi_1(V)$ and $\pi_2(V)$ are states of the \emph{first} and the \emph{second procedure invocation} and
we call the elements of their union $G_Q = \pi_1(V) \cup \pi_2(V)$ jointly the \emph{invocation states}.
The edges $E \subseteq V \times V$ represent pairs of \emph{similar transitions}, namely, an edge $((r_1, r_2), (s_1, s_2))$
indicates the existence of a transition pair
$\transition{r_1}{a,\eta_1/\eta_1}{s_1}$ and
$\transition{r_2}{a,\eta_2/\eta_2}{s_2}$ where $\eta_1, \eta_2 \in \{\bot, *\}$ (note that these transitions do not update the register).
We also require that each similarity graph satisfies the following assumptions:

\vspace*{0.25em}
\begin{adjustwidth}{3ex}{}
\begin{enumerate}
    \item[SG1]
    \label{SG:root}
     It must be \emph{rooted}, and we denote by $\mathit{root(G)}$ the root of a graph $G$,
    \item[SG2]
    \label{SG:no_register} each state from one invocation has to have exactly one similar state in the other invocation, i.e., $V$ is a bijection from $\pi_1(V)$ to $\pi_2(V)$,
    \item[SG3] states of each of the two invocations have the same signatures, i.e.,
    for any $r, s \in \pi_i(V),i\in\{1,2\}$, $\mathit{Sig}(r) = \mathit{Sig}(s)$.
    \item[SG4]
    \label{SG:disjoint}
    the signatures of the two invocations are disjoint, that is, for any $r \in \pi_1(V),s\in \pi_2(V)$, $\mathit{Sig}(r) \cap \mathit{Sig}(s) = \emptyset$.
\end{enumerate}
\end{adjustwidth}
These assumptions, together with the requirement that similar transitions do not update the register, will allow us to easily show that the introduction of a new procedure preserves the existing ones and does not break the nested structure (Lemma~\ref{lemma:correctness}).

\subsection{Procedure Creation}
\begin{wrapfigure}{r}{0.59\linewidth}
    \vspace*{-2.55em}
    \begin{algorithm}[H]
        \caption{createProcedure}
        \footnotesize
        \label{alg:mapProcedure}
        \DontPrintSemicolon
        \KwIn{SRA $A = (Q, \Sigma, \Gamma, \delta, q_0, F)$,\\
        \hspace*{\widthof{\KwIn{}}}\mbox{similarity graph of $A$ as $G = (V \subseteq Q \times Q, E)$}}
        \KwOut{SRA $B = (\algout{Q}, \Sigma, \algout{\Gamma}, \algout{\delta}, q_0, F)$}

        \vspace*{0.3em}
        \mbox{$B \gets A, Q_{proc} \gets \emptyset, (r_1, r_2) \gets root(V)$}\;
        \mbox{$\mathit{Id}(r_1) \gets \begin{cases} \{B.\mathit{newIdSymbol}()\} & \text{if } \mathit{Sig}(r_1) = \emptyset\\[-0.85mm] \mathit{Sig}(r_1) & \text{otherwise} \end{cases}$}\;
        \mbox{$\mathit{Id}(r_2) \gets \begin{cases} \{B.\mathit{newIdSymbol}()\} & \text{if } \mathit{Sig}(r_2) = \emptyset\\[-0.85mm] \mathit{Sig}(r_2) & \text{otherwise} \end{cases}$}\;
        $\algout{\Gamma} \gets \algout{\Gamma} \cup \mathit{Id}(r_1) \cup \mathit{Id}(r_2)$\;

        \ForAll{$(q_1, q_2) \in V$}
        {
            $q_{1,2} \gets B.\mathit{newState}()$\;
            $Q_{\mathit{proc}}(q_1) \gets q_{1,2}, Q_{\mathit{proc}}(q_2) \gets q_{1,2}$\;
            $\algout{Q} \gets \algout{Q} \cup \{q_{1,2}\}$\;
            $\mathit{Id}(q_1) \gets \mathit{Id}(r_1), \mathit{Id}(q_2) \gets \mathit{Id}(r_2)$\;
        }
        \mbox{$B \gets\mathit{createProcedureTransitions}(B, G, \mathit{Id}, Q_{\mathit{proc}})$}\;\vspace*{-1.1em}
        $B.\mathit{removeIncidentTransitions}(G_Q)$\;
        $B.\mathit{removeStates}(G_Q)$\;
        \Return{$B$}
        \normalsize
    \end{algorithm}
    \vspace*{-3em}
\end{wrapfigure}

In Algorithm \ref{alg:mapProcedure}, the procedure is constructed based on the similarity graph $G$ of the SRA $A$, and a new SRA with this procedure is returned.
To differentiate between the two original invocations, the algorithm assigns a unique \textit{invocation Id} to each.
Technically, the invocation Id is assigned to \emph{each state} of the invocation by the function $\mathit{Id} : Q \rightarrow 2^\Gamma$.
A new invocation Id is generated if the invocation does not belong to any procedure yet (i.e., if its signature is empty). Otherwise, the invocation reuses the existing signature of a state as its invocation Id. Here we rely on the assumption SG\ref{SG:disjoint} of the invocation signature disjointness that guarantees that the two invocations can indeed be distinguished by their signatures. The union of the two invocation Ids will later become the signature of the new procedure.
In lines \verb|2-3|, the invocation Id is assigned to the root states $r_1$ and $r_2$ of the similarity graph and then, in line \verb|9|, it is propagated to all states of the invocation.
In lines \verb|5-8|, the algorithm creates a new \textit{procedure state} $q_{1,2}$ for each pair of similar states $(q_1, q_2)$ from $V$. Which pair of invocation states corresponds to which procedure state is stored in the mapping $Q_{proc} : G_Q \rightarrow Q'$ to be later used by Algorithm \ref{alg:mapTransitions}, called on line \verb|10| to generate procedure transitions. In lines \verb|11-12|, the invocation states and their transitions that were used to create the procedure are removed from the automaton, as their functionality is now simulated by the procedure.

\subsection{Generating Procedure Transitions}
\vspace{-0.5em}
Algorithm \ref{alg:mapTransitions} creates procedure transitions.
Intuitively, we are replacing the original invocations, states $G_Q$, and incident transitions,
by a single procedure with the states $Q'$ and new transitions. The new transitions are supposed to simulate the original transitions of one invocation or the other depending on the register value, which is a symbol from one or the other invocation Id.
The rest of the section elaborates on the technical details of the implementation of this intuition.

Algorithm \ref{alg:mapTransitions} consists of five for-loops that generate five types of procedure transitions, each of them constructed from a different type of original invocation transition.
Lines \verb|2-7| generate new procedure transitions from \textit{invocation entry transitions}, lines \verb|8-16| from \textit{invocation exit transitions}, lines \verb|17-19| from \textit{invocation common transitions}, lines \verb|20-28| from \textit{invocation unique transitions}, and lines \verb|29-33| from \textit{invocation switch transitions}.
The function $pick$ is used to select an arbitrary element of a set.

\vspace*{-2em}

\begin{algorithmNoBottom}[!htp]
    \caption{createProcedureTransitions}
    \label{alg:mapTransitions}
    \footnotesize
    \DontPrintSemicolon
    \KwIn{SRA $A = (Q, \Sigma, \Gamma, \delta_{in}, I, F)$, similarity graph of $A$ as $G = (V \subseteq Q \times Q,E),$\\
    \hspace*{\widthof{\KwIn{}}}invocation state to invocation Id mapping $\mathit{Id} : G_Q \rightarrow 2^{\Gamma}$,\\
    \hspace*{\widthof{\KwIn{}}}invocation state to procedure state mapping $Q_{\mathit{proc}}: G_Q \rightarrow Q$}
    \KwOut{SRA $B = (Q, \Sigma, \Gamma, \algout{\delta}, I, F)$}
    \vspace*{0.1em}
    $B \gets A$\;
    \ForAll(\tcp*[f]{Paragraph Invocation Entry}){$r \crightarrow{$\{a, \alpha/\beta\}$} s \in \mathit{Entry}(G)$} {
        $s' \gets Q_{\mathit{proc}}(s)$\;
        \uIf(\tcp*[f]{new (Point 1)}){$\beta = \bot$} {
            $\beta' \gets \mathit{pick}(\mathit{Id}(s))$\;
            $\algout{\delta}.\mathit{add}(\transition{r}{a,\alpha/\beta'}{s})$\;

        }
        \textbf{else} {
            $\algout{\delta}.\mathit{add}(\transition{r}{a,\alpha/\beta}{s'})$\tcp*{copy (Point 2)}
        }
    }
    \ForAll(\tcp*[f]{Paragraph Invocation Exit}){$r \crightarrow{$\{a, \alpha/\beta\}$} s \in \mathit{Exit}(G)$} {
        $r' \gets Q_{\mathit{proc}}(r)$\;
        \uIf(\tcp*[f]{new (Point 1)}){$\alpha = \bot$} {
            $\alpha' \gets \mathit{pick}(\mathit{Id}(r))$\;
            $\algout{\delta}.\mathit{add}(\transition{r'}{a,\alpha'/\beta}{s})$\;
        } \uElseIf(\tcp*[f]{expand (Point 2)}){$\alpha = * \land \beta = *$} {
            \ForAll{$\eta \in \mathit{Id}(r)$} {
                $\algout{\delta}.\mathit{add}(\transition{r'}{a,\eta/\eta}{s})$\;
            }
        } \textbf{else} {
            $\algout{\delta}.\mathit{add}(\transition{r'}{a,\alpha/\beta}{s})$\tcp*{copy (Point 3)}
        }
    }

    \ForAll(\tcp*[f]{Paragraph Invocation Common}){$r \crightarrow{$\{a, \alpha/\beta\}$} s \in \mathit{Common}(G)$} {
        $r' \gets Q_{\mathit{proc}}(r), s' \gets Q_{\mathit{proc}}(s)$\;
        $\algout{\delta}.\mathit{add}(\transition{r'}{a,*/*}{s'})$\;
    }

    \ForAll(\tcp*[f]{Paragraph Invocation Unique}){$r \crightarrow{$\{a, \alpha/\beta\}$} s \in \mathit{Unique}(G)$} {
        $r' \gets Q_{\mathit{proc}}(r), s' \gets Q_{\mathit{proc}}(s)$\;
        \uIf(\tcp*[f]{new (Point 1)}){$\alpha = \bot \land \beta = \bot$} {
            $\eta \gets \mathit{pick}(\mathit{Id}(r))$\;
            $\algout{\delta}.\mathit{add}(\transition{r'}{a,\eta/\eta}{s'})$\;
        } \uElseIf(\tcp*[f]{expand (Point 2)}){$\alpha = * \land \beta = *$} {
            \ForAll{$\eta \in \mathit{Id}(r)$} {
                $\algout{\delta}.\mathit{add}(\transition{r'}{a,\eta/\eta}{s'})$\;
            }
        }
        \textbf{else} {
            $\algout{\delta}.\mathit{add}(\transition{r'}{a,\alpha/\beta}{s'})$\tcp*{copy (Point 3)}
        }
    }
\end{algorithmNoBottom}
\newpage
\begin{algorithmNoTop}[!htp]
    \vspace*{-0.25em}
    \DontPrintSemicolon
    \setcounter{AlgoLine}{28}
    \ForAll(\tcp*[f]{Paragraph Invocation Switch}){$r \crightarrow{$\{a, \alpha/\beta\}$} s \in \mathit{Switch}(G)$} {
        $r' \gets Q_{\mathit{proc}}(r), s' \gets Q_{\mathit{proc}}(s)$\;
        $\alpha' \gets \begin{cases}
                        \mathit{pick}(\mathit{Id}(r)) & \text{if } \alpha = \bot\hspace{5ex}\tcp*[f]{new}\\
                        \alpha & \text{otherwise}\hspace{3ex}\tcp*{copy}
                    \end{cases}$\;
        $\beta' \gets \begin{cases}
                        \mathit{pick}(\mathit{Id}(s)) & \text{if } \beta = \bot\hspace{5ex}\tcp*{new}\\
                        \beta & \text{otherwise}\hspace{3ex}\tcp*{copy}
                    \end{cases}$\;
        $\algout{\delta}.\mathit{add}(\transition{r'}{a,\alpha'/\beta'}{s'})$\;
    }
    \Return{$B$}
    \vspace*{-1.5em}
\end{algorithmNoTop}
\vspace*{-1em}
\paragraph{Invocation Entry.}
(\verb|lines 2-7|) Transitions of this type enter the invocations, and are formally defined by the set $\mathit{Entry}(G) = \{\transition{r}{a,\alpha/\beta}{s} \in \delta \mid r \notin G_Q \land s \in G_Q \}$.
We distinguish two cases:

\begin{enumerate}
    \item (\verb|lines 5-6|) If the transition's target state $s$ does not yet belong to any procedure, a~new symbol $\beta' \in \mathit{Id}(s)$ is assigned as a set symbol to the procedure transition. This symbol will be used to distinguish between invocations.
    \vspace*{0.25em}\item\begin{adjustbox}{minipage={\linewidth}, valign=t}
        \begin{wrapfigure}{tr}{0.5\linewidth}
            \vspace*{-2em}
            \begin{minipage}{\textwidth}
                \begin{subfigure}[!htp]{0.24\linewidth}
                    \resizebox{\textwidth}{!}{
                        \centering
                        \begin{tikzpicture}[node distance=1.9cm,every node/.style={font=\large}]
                            \node[state, initial] (q0) {$q_0$};
                            \node[state, above of=q0] (q1) {$q_1$};
                            \node[state, below of=q0] (q2) {$q_2$};
                            \node[state, thick, right of=q1] (q3) {$q_3$};
                            \node[state, thick, right of=q2] (q4) {$q_4$};
                            \node[state, thick, right of=q3] (q5) {$q_5$};
                            \node[state, thick, right of=q4] (q6) {$q_6$};
                            \node[state, accepting, below of=q5] (q7) {$q_7$};

                            \path[->, >=stealth']
                                    (q0) edge[left] node{$\begin{aligned}
                                                            x,\bot&/0\\[-2mm]
                                                            y,\bot&/1
                                                        \end{aligned}$} (q1)
                                    (q0) edge[left] node{$z,\bot/\bot$} (q2)
                                    (q0) edge[->, below right, dashed, pos=0.5] node{$m,\bot/1$} (q3)
                                    (q1) edge[above, dashed] node{$\begin{aligned}
                                                                        a,*&/*\\[-2mm]
                                                                        f,0&/1\\[-2mm]
                                                                        g,1&/1
                                                                    \end{aligned}$} (q3)
                                    (q2) edge[below, dashed] node{$e,\bot/\bot$} (q4)
                                    (q3) edge[above, thick] node{$b,*/*$} (q5)
                                    (q4) edge[below, thick] node{$b,\bot/\bot$} (q6)
                                    (q6) edge[right] node{$z,\bot/\bot$} (q7)
                                    (q5) edge[right] node{$\begin{aligned}
                                                                x,0&/\bot\\[-2mm]
                                                                y,1&/\bot
                                                        \end{aligned}$} (q7);
                        \end{tikzpicture}
                    }
                \end{subfigure}
                \begin{subfigure}[!htp]{0.24\linewidth}
                    \resizebox{\textwidth}{!}{
                        \centering
                        \begin{tikzpicture}[node distance=1.9cm, every node/.style={font=\large} ]
                            \node[state, initial] (q0) {$q_0$};
                            \node[state, above of=q0] (q1) {$q_1$};
                            \node[state, below of=q0] (q2) {$q_2$};
                            \node[rectangular state, right of=q0] (q34) {$q_{3,4}$};
                            \node[rectangular state, above right=1.2cm and 1.5cm of q34] (q56) {$q_{5,6}$};
                            \node[state, accepting, below of=q56] (f) {$f$};

                            \path[->, >=stealth']
                                    (q0) edge[left] node{$\begin{aligned}
                                                            x,\bot&/0\\[-2mm]
                                                            y,\bot&/1
                                                        \end{aligned}$} (q1)
                                    (q0) edge[above] node{$m,\bot/1$} (q34)
                                    (q0) edge[left] node{$z,\bot/\bot$} (q2)
                                    (q1) edge[out=east, in=north, above right, pos=0.1] node{$\begin{aligned}
                                                            a,*&/*\\[-2mm]
                                                            f,0&/1\\[-2mm]
                                                            g,1&/1
                                                        \end{aligned}$} (q34)
                                    (q2) edge[below right, pos=0.5] node{$e,\bot/2$} (q34)
                                    (q34) edge[above left, pos=0.8] node{$b,*/*$} (q56)
                                    (q56) edge[right] node{$\begin{aligned}
                                                                x,0&/\bot\\[-2mm]
                                                                y,1&/\bot\\[-2mm]
                                                                z,2&/\bot
                                                        \end{aligned}$} (f);
                        \end{tikzpicture}
                    }
                \end{subfigure}
            \end{minipage}
            \vspace*{-0.5em}
            \caption{Before and after the creation of procedure transitions from invocation entry transitions (dashed).}
            \label{fig:call_mapping}
            \vspace*{-2em}
        \end{wrapfigure}
        (\verb|line 7|) If the transition's target state $s$ already belongs to some procedure (i.e., the whole invocation entry transition belongs to some procedure), no changes have to be made to the register operations. The algorithm simply copies the transition to the corresponding procedure states specified by $Q_{proc}$. This is illustrated in Figure \ref{fig:call_mapping}, where the algorithm creates, from invocation transitions $\transition{q_1}{a,*/*}q_3$, $\transition{q_1}{f,0/1}{q_3}$, and $q_1\crightarrow{$\{g, 1 / 1\}$}q_3$, new procedure transitions $q_1\crightarrow{$\{a, */*\}$}q_{3,4}$, $q_1\crightarrow{$\{f, 0 / 1\}$}q_{3,4}$, and $q_1 \crightarrow{$\{g, 1 / 1\}$} q_{3,4}$.
    \end{adjustbox}
\end{enumerate}

\paragraph{Invocation Exit.}
(\verb|lines 8-16|) Transitions of this type leave invocations, formally defined by the set $\mathit{Exit}(G) = \{\transition{r}{a,\alpha/\beta}{s} \in \delta \mid r \in G_Q \land s \notin G_Q\}$.
We distinguish between three scenarios:
\begin{enumerate}
    \item (\verb|lines 11-12|) If the source state $r$ of the transition does not yet belong to any procedure, the algorithm assigns a new register symbol $\alpha' \in \mathit{Id}(r)$ as a~test symbol. This symbol ensures that the procedure return transition can only be taken with equivalent register value.
    \vspace*{0.25em}\item (\verb|lines 14-15|) If the invocation exit transition belongs to some procedure and does
    not use a~register (i.e., it is a simplification of transitions with register operations $\eta / \eta$ for all possible $\eta \in \mathit{Id}(r)$), the algorithm has to perform \textit{symbol expansion} by creating $|\mathit{Id}(r)|$ \textit{guard transitions} with register operations $\eta / \eta$ for each symbol $\eta \in \mathit{Id}(r)$ to ensure that this return transition will be taken only with the corresponding register value.
    The symbol expansion can be seen in Figure \ref{fig:return_mapping}, where
    $\mathit{Id}(q_3) = \{1, 2\}$ and the algorithm creates, from invocation exit transitions $q_3 \crightarrow{$\{b, */*\}$} q_5$ and $q_4 \crightarrow{$\{e, \bot/\bot\}$} q_6$, new procedure transitions that exit the procedure state $q_{3,4}$. To be able to correctly determine whether to go to the state $q_5$ or not, the algorithm has to

    \begin{adjustbox}{minipage={\linewidth}, valign=t}
        \begin{wrapfigure}{tr}{0.5\linewidth}
            \vspace*{-2.5em}
            \begin{minipage}{\textwidth}
                \begin{subfigure}[!htp]{0.24\linewidth}
                    \resizebox{\textwidth}{!}{
                        \centering
                        \begin{tikzpicture}[node distance=1.9cm, every node/.style={font=\large}]
                            \node[state, initial] (q0) {$q_0$};
                            \node[state, above of=q0] (q1) {$q_1$};
                            \node[state, below of=q0] (q2) {$q_2$};
                            \node[state, thick, right of=q1] (q3) {$q_3$};
                            \node[state, thick, right of=q2] (q4) {$q_4$};
                            \node[state, thick, right of=q3] (q5) {$q_5$};
                            \node[state, thick, right of=q4] (q6) {$q_6$};
                            \node[state, accepting, below of=q5] (q7) {$q_7$};

                            \path[->, >=stealth']
                                    (q0) edge[left] node{$\begin{aligned}
                                                            x,\bot&/0\\[-1mm]
                                                            y,\bot&/1
                                                        \end{aligned}$} (q1)
                                    (q0) edge[left] node{$z,\bot/\bot$} (q2)
                                    (q1) edge[above, thick] node{$a,*/*$} (q3)
                                    (q2) edge[below, thick] node{$a,\bot/\bot$} (q4)
                                    (q3) edge[above, dashed] node{$b,*/*$} (q5)
                                    (q3) edge[below left, dashed] node{$m,0/\bot$} (q7)
                                    (q4) edge[below, dashed] node{$e,\bot/\bot$} (q6)
                                    (q6) edge[right] node{$z,\bot/\bot$} (q7)
                                    (q5) edge[right] node{$\begin{aligned}
                                                                x,0&/\bot\\[-2mm]
                                                                y,1&/\bot
                                                        \end{aligned}$} (q7);
                        \end{tikzpicture}
                    }
                \end{subfigure}
                \begin{subfigure}[!htp]{0.24\linewidth}
                    \resizebox{\textwidth}{!}{
                        \centering
                        \begin{tikzpicture}[node distance=1.9cm, every node/.style={font=\large}]
                            \node[state, initial] (q0) {$q_0$};
                            \node[rectangular state, thick, above of=q0] (q12) {$q_{1,2}$};
                            \node[rectangular state, thick, right of=q0] (q34) {$q_{3,4}$};
                            \node[state, above right=1.2cm and 1.5cm of q34] (q5) {$q_{5}$};
                            \node[state, accepting, below of=q5] (f) {$f$};
                            \node[state, below of=f] (q6) {$q_6$};

                            \path[->, >=stealth']
                                    (q0) edge[left, pos=0.45] node{$\begin{aligned}
                                                            x,\bot&/0\\[-2mm]
                                                            y,\bot&/1\\[-2mm]
                                                            z,\bot&/2
                                                        \end{aligned}$} (q12)
                                    (q12) edge[above right, pos=0.01, thick, out=east, in=north west] node{$a,*/*$} (q34)
                                    (q34) edge[above left, pos=0.99, out=north east, in=west] node{$\begin{aligned}
                                                                b,0&/0\\[-2mm]
                                                                b,1&/1
                                                        \end{aligned}$} (q5)
                                    (q34) edge[above] node{$m,0/\bot$} (q7)
                                    (q34) edge[below left, pos=0.125, out=south, in=west] node{$e,2/\bot$} (q6)
                                    (q6) edge[right] node{$z,\bot/\bot$} (f)
                                    (q5) edge[right, pos=0.45] node{$\begin{aligned}
                                                                x,0&/\bot\\[-2mm]
                                                                y,1&/\bot
                                                        \end{aligned}$} (f);

                        \end{tikzpicture}
                    }
                \end{subfigure}
            \end{minipage}
            \vspace*{-0.25em}
            \caption{Before and after the creation of procedure transitions from invocation exit transitions (dashed).}
            \label{fig:return_mapping}
            \vspace*{-2em}
        \end{wrapfigure}
        perform symbol expansion of $\mathit{Id}(q_3)$ by creating guard transitions with equivalent test and set symbols for each register symbol from $\mathit{Id}(q_3)$.
        This results in the creation of guard transitions $q_{3,4} \crightarrow{$\{b, 0 / 0\}$} q_5$~and $q_{3,4}\crightarrow{$\{b, 1 / 1\}$}q_5$. Note that the symbol expansion increases the
        number of transitions in the automaton. Because of that, we will prioritize similarity graphs that require fewer symbol expansions (it also happens with invocation unique transitions discussed below).

    \end{adjustbox}

    \vspace*{0.35em}\item (\verb|line 16|) If the transition has set symbol different from $\bot$ and $*$, the transition is simply copied to the corresponding procedure state specified by $Q_{\mathit{proc}}$.
\end{enumerate}

\paragraph{Invocation Common.}
(\verb|lines 17-19|)
Transitions of this type are shared by the two invocations, formally defined by the set
$\mathit{Common}(G) = \{\transition{r}{a,\eta/\eta}{s} \in \delta \mid r, s \in G_Q \land \eta \in \{\bot, *\} \land \exists r', s' \in G_Q : \transition{r'}{a,\eta'/\eta'}{s'} \in \delta \land \eta' \in \{\bot, *\} \land (((r, r'), (s, s')) \in E \lor ((r', r),$ $(s', s)) \in E)\}$.
Since they exist in both invocations, there is no need to use register sym-
\begin{wrapfigure}{r}{0.5\linewidth}
    \vspace*{-2em}
    \begin{minipage}{\textwidth}
        \begin{subfigure}[!htp]{0.24\linewidth}
            \resizebox{\textwidth}{!}{
                \centering
                \begin{tikzpicture}[node distance=1.7cm]
                    \node[state, initial] (q0) {$q_0$};
                    \node[state, thick, above of=q0] (q1) {$q_1$};
                    \node[state, thick, below of=q0] (q2) {$q_2$};
                    \node[state, thick, right of=q1] (q3) {$q_3$};
                    \node[state, thick, right of=q2] (q4) {$q_4$};
                    \node[state, thick, right of=q3] (q5) {$q_5$};
                    \node[state, thick, right of=q4] (q6) {$q_6$};
                    \node[state, accepting, below of=q5] (q7) {$q_7$};

                    \path[->, >=stealth']
                            (q0) edge[left] node{$\begin{aligned}
                                                    x,\bot&/0\\[-2mm]
                                                    y,\bot&/1
                                                \end{aligned}$} (q1)
                            (q0) edge[left] node{$z,\bot/\bot$} (q2)
                            (q1) edge[above, thick, dashed] node{$a,*/*$} (q3)
                            (q1) edge[below, bend right] node{$e,1/1$} (q5)
                            (q2) edge[below, thick, dashed] node{$a,\bot/\bot$} (q4)
                            (q3) edge[above, thick, dashed] node{$b,*/*$} (q5)
                            (q4) edge[below, thick, dashed] node{$b,\bot/\bot$} (q6)
                            (q6) edge[right] node{$z,\bot/\bot$} (q7)
                            (q5) edge[right] node{$\begin{aligned}
                                                        x,0&/\bot\\[-2mm]
                                                        y,1&/\bot
                                                \end{aligned}$} (q7);
                \end{tikzpicture}
            }
        \end{subfigure}
        \begin{subfigure}[!htp]{0.24\linewidth}
            \resizebox{\textwidth}{!}{
                \centering
                \begin{tikzpicture}[node distance=1.7cm]
                    \node[state, initial] (q0) {$q_0$};
                    \node[rectangular state, above of=q0] (q12) {$q_{1,2}$};
                    \node[rectangular state, right of=q12] (q34) {$q_{3,4}$};
                    \node[rectangular state, right of=q34] (q56) {$q_{5,6}$};
                    \node[state, accepting, below of=q56] (q7) {$q_7$};

                    \path[->, >=stealth']
                            (q0) edge[left] node{$\begin{aligned}
                                                    x,\bot&/0\\[-2mm]
                                                    y,\bot&/1\\[-2mm]
                                                    z,\bot&/2
                                                \end{aligned}$} (q12)
                            (q12) edge[above] node{$a,*/*$} (q34)
                            (q12) edge[below, bend right] node{$e,1/1$} (q56)
                            (q34) edge[above] node{$b,*/*$} (q56)
                            (q56) edge[right] node{$\begin{aligned}
                                                        x,0&/\bot\\[-2mm]
                                                        y,1&/\bot\\[-2mm]
                                                        z,2&/\bot
                                                \end{aligned}$} (q7);
                \end{tikzpicture}
            }
        \end{subfigure}
    \end{minipage}
    \vspace*{-0.25em}
    \caption{Before and after the creation of procedure transitions from invocation common transitions (dashed).}
    \label{fig:inner_mapping}
    \vspace*{-2em}
\end{wrapfigure}
bols to determine if the procedure transition should be taken. Consequently, the transition is simply copied to the procedure. It is important to note that it is only the presence of invocation common transitions that causes the reduction in the number of transitions after introducing the procedure. That is why we will prioritize similarity graphs that contain many invocation common transitions.

\paragraph{Invocation Unique.}
(\verb|lines 20-28|) Transitions of this type are unique to only one of the invocations, formally defined as $\mathit{Unique}(G) = \{ \transition{r}{a,\alpha/\beta}{s} \in  \delta \mid r, s \in \pi_1(V) \Leftrightarrow r, s \notin \pi_2(V) \} \setminus \mathit{Common}(G)$. Procedure transitions, constructed on the basis of these invocation unique transitions, must be guarded by testing the register value. The construction of procedure transitions from invocation unique transitions is similar to that from invocation entry transitions:
\begin{enumerate}
    \item (\verb|lines 23-24|) If the transition does not yet belong to any procedure, the algorithm uses a new symbol $\eta \in \mathit{Id}(r)$ for test and set operation to guard the transition. This ensures that the transition will be taken only with the corresponding register value.
    \vspace*{0.25em}\item (\verb|lines 26-27|) If both the source state $r$ and the target state $s$ belong to some procedure and the transition does not use a register (i.e., it is a simplification of transitions with register operations $\eta / \eta$ for each possible $\eta \in \mathit{Id}(r)$), the algorithm performs symbol expansion by creating $|\mathit{Id}(r)|$ guard transitions with register operations $\eta / \eta$ for each symbol $\eta \in \mathit{Id}(r)$. The symbol expansion is equivalent to that for invocation exit transitions. The example of this symbol expansion can be \mbox{seen in Figure \ref{fig:unique_mapping}, where $\mathit{Id}(q_1) = \{1, 2\}$ and the algorithm creates, from invocation}
    \newpage
    \begin{adjustbox}{minipage={\linewidth}, valign=t}

        \begin{wrapfigure}{r}{0.5\linewidth}
            \vspace*{-2.25em}
            \begin{minipage}{\textwidth}
                \begin{subfigure}[!htp]{0.24\linewidth}
                    \resizebox{\textwidth}{!}{
                        \centering
                        \begin{tikzpicture}[node distance=1.7cm]
                            \node[state, initial] (q0) {$q_0$};
                            \node[state, thick, above of=q0] (q1) {$q_1$};
                            \node[state, thick, below of=q0] (q2) {$q_2$};
                            \node[state, thick, right of=q1] (q3) {$q_3$};
                            \node[state, thick, right of=q2] (q4) {$q_4$};
                            \node[state, thick, right of=q3] (q5) {$q_5$};
                            \node[state, thick, right of=q4] (q6) {$q_6$};
                            \node[state, accepting, below of=q5] (q7) {$q_7$};

                            \path[->]
                                    (q0) edge[left] node{$\begin{aligned}
                                                            x,\bot&/0\\[-2mm]
                                                            y,\bot&/1
                                                        \end{aligned}$} (q1)
                                    (q0) edge[left] node{$z,\bot/\bot$} (q2)
                                    (q1) edge[above, thick] node{$a,*/*$} (q3)
                                    (q1) edge[above, bend left=90, dashed] node{$f,*/*$} (q3)
                                    (q1) edge[below, dashed, bend right] node{$e,1/1$} (q5)
                                    (q2) edge[below, thick] node{$a,\bot/\bot$} (q4)
                                    (q3) edge[above, thick] node{$b,*/*$} (q5)
                                    (q4) edge[below, thick] node{$b,\bot/\bot$} (q6)
                                    (q4) edge[above, out=45, in=135, dashed] node{$g,\bot/\bot$} (q6)
                                    (q6) edge[right] node{$z,\bot/\bot$} (q7)
                                    (q5) edge[right] node{$\begin{aligned}
                                                                x,0&/\bot\\[-2mm]
                                                                y,1&/\bot
                                                        \end{aligned}$} (q7);
                        \end{tikzpicture}
                    }
                \end{subfigure}
                \begin{subfigure}[!htp]{0.24\linewidth}
                    \resizebox{\textwidth}{!}{
                        \centering
                        \begin{tikzpicture}[node distance=1.7cm]
                            \node[state, initial] (q0) {$q_0$};
                            \node[rectangular state, above of=q0] (q12) {$q_{1,2}$};
                            \node[rectangular state, right of=q12] (q34) {$q_{3,4}$};
                            \node[rectangular state, right of=q34] (q56) {$q_{5,6}$};
                            \node[state, accepting, below of=q56] (q7) {$q_7$};

                            \path[->]
                                    (q0) edge[left] node{$\begin{aligned}
                                                            x,\bot&/0\\[-2mm]
                                                            y,\bot&/1\\[-2mm]
                                                            z,\bot&/2
                                                        \end{aligned}$} (q12)
                                    (q12) edge[above] node{$\begin{aligned}
                                                                a,*&/*\\[-2mm]
                                                                f,0&/0\\[-2mm]
                                                                f,1&/1
                                                        \end{aligned}$} (q34)
                                    (q12) edge[below, bend right] node{$e,1/1$} (q56)
                                    (q34) edge[above] node{$\begin{aligned}
                                                                b,*&/*\\[-2mm]
                                                                g,2&/2
                                                        \end{aligned}$} (q56)
                                    (q56) edge[right] node{$\begin{aligned}
                                                                x,0&/\bot\\[-2mm]
                                                                y,1&/\bot\\[-2mm]
                                                                z,2&/\bot
                                                        \end{aligned}$} (q7);
                        \end{tikzpicture}
                    }
                \end{subfigure}
            \end{minipage}
            \vspace*{-0.4em}
            \caption{Before and after the creation of procedure transitions from invocation unique transitions (dashed).}
            \label{fig:unique_mapping}
            \vspace*{-2em}
        \end{wrapfigure}
        unique transition $q_1 \crightarrow{$\{f, */*\}$} q_3$, new procedure transition between procedure states $q_{1,2}$ and $q_{3,4}$. To prohibit taking this transition with register symbol 2, the algorithm has to create guard transitions with equivalent test and set symbols for each symbol that can appear in the register while in the invocation state $q_1$. This results in the creation of guard transitions $q_{1,2} \crightarrow{$\{f, 0 / 0\}$} q_{3,4}$ and $q_{1,2} \crightarrow{$\{f, 1 / 1\}$} q_{3,4}$ and increase in the number of transitions in the automaton.
        \vspace{-0.8em}
    \end{adjustbox}

    \vspace*{0.3em}\item (\verb|line 28|) If the invocation transition has test and set symbols different from~$\bot$ and~$*$, the transition is copied to procedure state specified by $Q_{\mathit{proc}}$.
\end{enumerate}

\paragraph{Invocation Switch.}
\begin{wrapfigure}{r}{0.5\linewidth}
    \vspace*{-3.1em}
    \begin{minipage}{\textwidth}
        \begin{subfigure}[!htp]{0.24\linewidth}
            \resizebox{\textwidth}{!}{
                \centering
                \begin{tikzpicture}[node distance=1.7cm]
                    \node[state, initial] (q0) {$q_0$};
                    \node[state, thick, above of=q0] (q1) {$q_1$};
                    \node[state, thick, below of=q0] (q2) {$q_2$};
                    \node[state, thick, right of=q1] (q3) {$q_3$};
                    \node[state, thick, right of=q2] (q4) {$q_4$};
                    \node[state, thick, right of=q3] (q5) {$q_5$};
                    \node[state, thick, right of=q4] (q6) {$q_6$};
                    \node[state, accepting, below of=q5] (q7) {$q_7$};

                    \path[->]
                            (q0) edge[left] node{$\begin{aligned}
                                                    x,\bot&/0\\[-2mm]
                                                    y,\bot&/1
                                                \end{aligned}$} (q1)
                            (q0) edge[left] node{$z,\bot/\bot$} (q2)
                            (q1) edge[above, thick] node{$a,*/*$} (q3)
                            (q1) edge[above, bend left=90, dashed] node{$f,0/1$} (q3)
                            (q2) edge[below, thick] node{$a,\bot/\bot$} (q4)
                            (q3) edge[above, thick] node{$b,*/*$} (q5)
                            (q3) edge[left, bend right=10, dashed] node{$m,1/\bot$} (q4)
                            (q4) edge[below, thick] node{$b,\bot/\bot$} (q6)
                            (q4) edge[right, bend right=10, dashed] node{$n,\bot/0$} (q3)
                            (q6) edge[right] node{$z,\bot/\bot$} (q7)
                            (q5) edge[right] node{$\begin{aligned}
                                                        x,0&/\bot\\[-2mm]
                                                        y,1&/\bot
                                                \end{aligned}$} (q7);
                \end{tikzpicture}
            }
        \end{subfigure}
        \begin{subfigure}[!htp]{0.24\linewidth}
            \resizebox{\textwidth}{!}{
                \centering
                \begin{tikzpicture}[node distance=1.7cm]
                    \node[state, initial] (q0) {$q_0$};
                    \node[rectangular state, above of=q0] (q12) {$q_{1,2}$};
                    \node[rectangular state, right of=q12] (q34) {$q_{3,4}$};
                    \node[rectangular state, right of=q34] (q56) {$q_{5,6}$};
                    \node[state, accepting, below of=q56] (q7) {$q_7$};

                    \path[->]
                            (q0) edge[left, pos=0.45] node{$\begin{aligned}
                                                    x,\bot&/0\\[-2mm]
                                                    y,\bot&/1\\[-2mm]
                                                    z,\bot&/2
                                                \end{aligned}$} (q12)
                            (q12) edge[above] node{$\begin{aligned}
                                                        f,0&/1\\[-2mm]
                                                        a,*&/*
                                                \end{aligned}$} (q34)
                            (q34) edge[above] node{$b,*/*$} (q56)
                            (q34) edge[loop below] node{$\begin{aligned}
                                                        m,1&/2\\[-2mm]
                                                        n,2&/0
                                                    \end{aligned}$} (q34)
                            (q56) edge[right] node{$\begin{aligned}
                                                        x,0&/\bot\\[-2mm]
                                                        y,1&/\bot\\[-2mm]
                                                        z,2&/\bot
                                                \end{aligned}$} (q7);
                \end{tikzpicture}
            }
        \end{subfigure}
    \end{minipage}
    \vspace*{-0.25em}
    \caption{Before and after the creation of procedure transitions from invocation switch switch transitions (dashed).}
    \label{fig:switch_mapping}
    \vspace*{-3em}
\end{wrapfigure}
(\verb|lines 29-33|) Transitions of this type lead from states of the first invocation to states of the second invocation or vice versa. The set of these transitions is defined as $\mathit{Switch}(G) = \{\transition{r}{a,\alpha/\beta}{s'} \in \delta \,|\, r\in\pi_1(V) \Leftrightarrow s' \in \pi_2(V)\}$. Because of the change between invocations, the register value must change accordingly.

\vspace*{1em}
The following lemma summarizes the correctness of the procedure creation: it preserves the language and keeps the well nested hierarchical structure of procedures.

\begin{lemma}
Let $A$ be an SRA with nested procedures, and let $B$ is the SRA returned by Algorithm~\ref{alg:mapProcedure}.
Then $L(A) = L(B)$ and $B$ has nested procedures.
\label{lemma:correctness}
\end{lemma}

\section{Finding Procedures}
\vspace{-0.5em}
The reduction by procedure finding works by iteratively identifying similarity graphs and introducing new procedures.
In every iteration, we search for a similarity graph such that the introduction of the procedure will decrease the size of the automaton as much as possible. The decrease in automaton size is called the \emph{gain} of the similarity graph.
In this work, we choose to measure the size of the automaton as the number of its transitions.
The number of transitions usually determines the size of the automata representation,
and it particularly determines the resources needed to store the automaton on an FPGA chip in the works \cite{ApproxRed,FPGA_based_network_scaning},
which motivates our experiments.
Our algorithms could easily be adapted to different measures of the automata size, e.g., ones where the number of states is more important (though, our experiments show that our reduction driven by the number of transitions does also lead to a significant reduction of the number of states).

\subsection{Procedure Gain}
\vspace{-0.25em}
\label{sec:procedure_gain}
Let us now discuss the notation of gain $\mathit{Gain}(G)  \in \mathbb{Z}$ of the similarity graph  $G = (V, E)$. The gain gives a measure of how creating a procedure based on the similarity graph impacts the size of the automaton.
It can be determined by inspecting the transitions of the similarity graph.
The only positive component of the gain is the \textit{common transitions gain}, $\mathit{Gain}_{\mathit{com}}(G)$,
defined as the number of invocation common transitions divided by 2.
The gain is then negatively affected by the \textit{unique transitions loss}, $\mathit{Loss}(\mathit{Unique}(G))$, which represents the increase in the number of transitions caused by the creation of new guard transitions during symbol expansion, as described in lines \verb|26-27| of Algorithm \ref{alg:mapTransitions}, for the invocation unique transitions. Another negative factor similar to unique transitions loss is \textit{exit transitions loss}, $\mathit{Loss}(\mathit{Exit}(G))$, which represents the increase in the number of transitions caused by the creation of guard transitions during symbol expansion, as described in lines \verb|14-15| of Algorithm \ref{alg:mapTransitions}, for the invocation exit transitions.
The following lemma specifies how the exact gain is computed from the similarity graph $G$.

\begin{lemma}
    For a similarity graph $G$, we define $\mathit{Gain}(G)$ as $\mathit{Gain}(G) = \mathit{Gain}_{\mathit{com}}(G) - \mathit{Loss}(\mathit{Unique}(G)) - \mathit{Loss}(\mathit{Exit}(G))$ where:
    $$
    \mathit{Gain}_{\mathit{com}}(G) = \nicefrac{1}{2}{|\mathit{Common}(G)|} \hspace{7ex}
    \mathit{Loss}(\delta') = \hspace*{-4ex}\sum_{\,\transition{r}{a,*/*}{s} \in \delta'}\hspace{-4ex} |\mathit{Sig}(r)| - 1
    $$
\end{lemma}
The intuition behind the formula for $\mathit{Loss}(\delta')$ in the lemma is that
each unique and exit transition must be replaced by $|\mathit{Sig}(r)|$ guard transitions, one for each symbol from $\mathit{Sig}(r)$. This increases the number of transitions by $|\mathit{Sig}(r)| - 1$.

\subsection{Finding Similarity Graphs with a Large Gain}
\vspace{-0.25em}
\label{sec:on_similarity_finding}

Every similarity graph is a sub-graph of the \emph{self-product} of the SRA $A = (Q, \Sigma, \Gamma, \delta,\break I, F)$. The self-product is represented by the directed graph $G(A) = (V_{G}, E_{G})$ where:
\begin{enumerate}
    \item $V_{G} = ( Q \setminus (I \cup F) )^2$ and
    \item $((r_1, r_2), (s_1, s_2)) \in E_{G}$ iff for some $a \in \Sigma$, $\transition{r_1}{a,\eta_1/\eta_1}{s_1} \in \delta$ and  $\transition{r_2}{a,\eta_2/\eta_2}{s_2} \in \delta$, where $\eta_1, \eta_2 \in \{\bot, *\}$.
\end{enumerate}
An exhaustive exploration of sub-graphs of the self-product, which is already qua\-dra\-tic to the size of $A$, would not be feasible, hence
we a-priori narrow the class of similarity graphs to those that are \emph{linear}, meaning that their spanning tree is a simple path. Procedures of this form occur often in our benchmarks.

We then enumerate all vertices of the self-product and explore the space of similarity graphs starting from them, and select the one with the best gain.
Since it would still not be practically feasible to explore all linear graphs starting at every node of the self-product (as there is an exponential number of them), we utilize the following heuristics.

First, from every node $n$ of the self-product, we explore all linear graphs until the pre-set depth $d$, which we call the \emph{depth limit}, and assign to $n$ the maximum gain $\mathit{Gain}^d(n)$ produced by these graphs.
We then generate the desired linear similarity graph $G$ by a~greedy search guided by the function $\mathit{Gain}^d$. We initialize $G$ with its root node, the node with the highest $\mathit{Gain}^d$ in the self-product, and keep iteratively extending the path from the root. In every iteration, we add an edge towards the node with the highest possible $\mathit{Gain}^d$.
The extended path/similarity graph must also satisfy the conditions SG1-4 from the definition of a similarity graph, and the construction is terminated when the path cannot be extended anymore in a way that increases its gain.
We will denote the described procedure by $\mathit{findSimGraph}$ and $\mathit{findSimGraph}(A,d)$ is the similarity graph obtained by running it on the SRA $A$ with the depth limit $d$.

\paragraph{Impact of the Depth Limit Parameter $d$.}
One might wonder how the depth limit $d$ impacts the quality of the similarity graph and the resulting reduction. This also raises the question of how to set an appropriate depth limit. To illustrate the importance of choosing the right depth limit, consider the example in Figure \ref{fig:depth_limit}.
\vspace*{-1.5em}
\begin{figure}
    \centering
    \begin{subfigure}[!htp]{0.30\linewidth}
        \resizebox{\textwidth}{!}{
            \centering
            \begin{tikzpicture}[node distance=1.7cm]
                \node[state, initial] (q0) {$q_0$};
                \node[state, right of=q0] (q1) {$q_1$};
                \node[state, above of=q1] (q2) {$q_2$};
                \node[state, accepting, right of=q1] (q3) {$q_3$};

                \path[->]
                    (q0) edge[below] node{$x, \bot/0$} (q1)
                    (q1) edge[right, bend right=20] node{$a, */*$} (q2)
                    (q2) edge[left, bend right=20] node{$\begin{aligned}
                                                    a, 0&/1\\[-3.5mm]
                                                    \scalebox{0.8}{\vdots}&\\[-2mm]
                                                    a, 49&/50
                                                \end{aligned}$} (q1)
                    (q1) edge[below] node{$y, 50/\bot$} (q3);
            \end{tikzpicture}
        }
        \subcaption*{$d = 1$}
        \vspace*{-1mm}
    \end{subfigure}
    \begin{subfigure}[!htp]{0.34\linewidth}
        \resizebox{\textwidth}{!}{
            \centering
            \begin{tikzpicture}[node distance=1.7cm]
                \node[state, initial] (q0) {$q_0$};
                \node[state, right of=q0] (q1) {$q_1$};
                \node[state, above of=q1] (q3) {$q_3$};
                \node[state, accepting, right of=q1] (q4) {$q_4$};
                \node[state, above of=q4] (q2) {$q_2$};

                \path[->]
                    (q0) edge[below] node{$x, \bot/0$} (q1)
                    (q1) edge[left, pos=0.7] node{$a, */*$} (q2)
                    (q2) edge[above] node{$a, */*$} (q3)
                    (q3) edge[left, pos=0.4] node{$\begin{aligned}
                                                a, 0&/1\\[-3.5mm]
                                                \scalebox{0.8}{\vdots}&\\[-2mm]
                                                a, 32&/33
                                            \end{aligned}$} (q1)
                    (q2) edge[right] node{$y, 33/\bot$} (q4);
            \end{tikzpicture}
        }
        \subcaption*{$d = 2$}
    \end{subfigure}
    \begin{subfigure}[!htp]{0.3\linewidth}
        \resizebox{\textwidth}{!}{
            \vspace{-2em}
            \centering
            \begin{tikzpicture}[node distance=1.7cm]
                \node[state, initial] (q0) {$q_0$};
                \node[state, above of=q0] (q1) {$q_1$};
                \node[right=0.25cm of q1] (aux) {$\cdots$};
                \node[state, right=0.25cm of aux] (q50) {$q_{50}\hspace{-0.2ex}$};
                \node[state, below of=q50] (q51) {$q_{51}\hspace{-0.2ex}$};
                \node[state, accepting, right of=q51] (q52) {$q_{52}\hspace{-0.2ex}$};

                \draw[decorate,decoration={brace, raise=0.15cm}] (q1) -- (q50) node[midway, above=0.3cm]{\footnotesize $a,*/*$ 49 times};

                \path[->]
                    (q0) edge[left] node{$x, \bot/0$} (q1)
                    (q1) edge[above] node{} (aux)
                    (aux) edge[above] node{} (q50)
                    (q50) edge[below, bend left=20] node{$a, 0/1$} (q1)
                    (q50) edge[right] node{$a, 1/\bot$} (q51)
                    (q51) edge[below] node{$y, \bot/\bot$} (q52);
                \end{tikzpicture}
            }
            \vspace{-4mm}
            \subcaption*{$d = 49$}
            \vspace{1.5mm}
        \end{subfigure}
    \vspace*{-1em}
    \caption{An example illustrating the effect of the depth limit $d$ on the resulting SRA. The SRAs are constructed from an NFA representing a word $xa_1a_2\cdots a_{100}y$.}
    \label{fig:depth_limit}
\end{figure}
\vspace*{-1.25em}

The example demonstrates the construction of SRAs with $d = 1, 2,$ and $49$ from an NFA representing the word $xa_1a_2\cdots a_{100}y$. Since the depth determines how far ahead the heuristic looks when building the similarity graph, the depth limit $d$ can, in the worst case, lead to the creation of procedures of maximum length $d$. This happens when the algorithm $\mathit{findSimGraph}$ begins constructing the similarity graph from the end of the repeating sequence.

Clearly, as shown in Table \ref{tab:depth_limit}, setting the depth limit too low leads to a suboptimal reduction. Conversely, an excessively large depth limit not only slows down computation
\begin{wrapfigure}{r}{0.5\linewidth}
    \vspace*{-2.35em}
    \makeatletter\def\@captype{table}\makeatother
    \scriptsize
    \centering
    \setlength{\tabcolsep}{2.5pt}
    \renewcommand{\arraystretch}{1.1}
    \begin{minipage}{0.15\textwidth}
        \begin{tabular}{|c||c|c|c|}
            \hline
            $d$ & $Q$ & $\Gamma$ & $\delta$ \\
            \hline\hline
            1 & 4 & 51 & 53 \\
            2 & 5 & 34 & 37 \\
            49 & 53 & 2 & 53 \\
            \hline
        \end{tabular}
    \end{minipage}
    \hfill
    \begin{minipage}{0.33\textwidth}
        \vspace*{1.5em}
        \caption{Impact of the depth limit $d$ on the size of the resulting SRAs in Figure \ref{fig:depth_limit}.}
        \label{tab:depth_limit}
    \end{minipage}
    \vspace{-4em}
\end{wrapfigure}
but may also result in suboptimal reduction. The optimal depth limit varies depending on the structure of the NFA being reduced. By minimizing the number of transitions and the sum of the number of states and register symbols, we found that the best depth limit for this example is $d = 9$. Based on this observation and further experiments, we set the default depth limit in our benchmarks to 10.

\section{The Main Size Reducing Algorithm and Post-Processing}\label{sec:post_processing}
\vspace{-0.25em}

\begin{wrapfigure}{r}{0.52\textwidth}
    \vspace*{-2.4em}
    \begin{algorithm}[H]
        \DontPrintSemicolon
        \KwIn{SRA $A{=}(Q, \Sigma, \Gamma, \delta, I, F)$, \mbox{depth limit $d \in \mathbb{N}_0$}}
        \KwOut{\mbox{smaller $A$ with new procedures}}
        \vspace*{0.3em}

        $G \gets \mathit{findSimGraph}(A,d)$\;
        \While{$\mathit{Gain}(G) > 0$}
        {
            $A \gets \mathit{createProcedure}(A, G)$\;
            $G \gets \mathit{findSimGraph}(A,d)$\;
        }
        \Return{$\textit{post-process}(A)$}
        \caption{\mbox{Main Reduction Algorithm}}
        \label{alg:main}
    \end{algorithm}
    \vspace*{-2em}
\end{wrapfigure}
Our method for reducing the size of a given SRA $A$ (possibly converted from an NFA) is summarized in the simple pseudocode of Algorithm~\ref{alg:main}. The function $\mathit{createProcedure}$
is run repeatedly on $A$ with the similarity graph found by the heuristic $\mathit{findSimGraph}$ from Section~\ref{sec:on_similarity_finding} while the heuristic keeps finding similarity graphs with a positive gain. The resulting SRA is then post-processed by reducing the size of the register alphabet and by removing vacuous guards of transitions. The post-processing, described below, has a significant effect, especially on the size of the register alphabet.

\paragraph{Merging Register Symbols.}
With multiple procedures created, the register alphabet often contains many symbols that cannot \textit{meet} (i.e., appear in the register while in the same state). Consequently, it is not necessary to keep these symbols distinct.
To identify register symbols that can appear in the register while in the same state, we establish an equivalence relation $\sim$ on the register alphabet $\Gamma$ such that $\alpha \sim \beta  \iff \exists q \in Q : \alpha \in \mathit{Sig}(q) \land \beta \in \mathit{Sig}(q)$.
The size of the largest equivalence class $C$ of $\sim$ determines the number of needed register symbols.
We will thus use only the symbols from the largest class $C$ and rename the rest of the symbols to symbols of this class as follows. For every equivalence class $D$ other than $C$, we construct an injective mapping $f:D\rightarrow C$, and replace all occurrences of \mbox{every $\alpha\in D$ in the automaton by $f(\alpha)$.}

\vspace{-0.5em}
\paragraph{Removing Vacuous Guards.}
A set $\{\transition{r}{a,\eta/\eta}{s} \subseteq \delta \mid \eta \in \mathit{Sig}(r)\}$ of \emph{vacuous guard transitions} between states $r$ and $s$ over an input symbol $a$, i.e, a set that 1) contains transitions that do not modify the register and 2) contains a transition for each possible register symbol $\eta \in \mathit{Sig}(r)$, can be collapsed into a single transition $\transition{r}{a,*/*}{s}$. Indeed, a test that always returns true can be removed.

\vspace{-0.25em}
\section{Experimental Evaluation}
\vspace{-0.25em}
We evaluated our approach on two groups of benchmarks. The first consists of automata derived from filtering rules of Snort network intrusion detection system \cite{Snort}.
This benchmark is highly practically relevant.
In high-speed networks operating at 10 Gbps or higher, the scanning must be accelerated on hardware, using e.g. FPGA chips  \cite{FPGA_based_network_scaning,ApproxRed},
which have very limited memory and where the needed speed is achieved by running copies of the same automaton in parallel.
The automata size hence becomes a critical bottleneck.
The second benchmark consists of automata from several applications, namely, abstract regular model checking \cite{ARMC}, from string constraint solver Z3-Noodler \cite{z3-noodler-git}, and from regular expressions used in email validation collected from the RegexLib library \cite{RegexLib} and parametric regular expressions experimented with in \cite{TACAS13}.

For both benchmarks,
the automata that our method got on the input were already reduced by the tool RABIT/Reduce (version~2.5) \cite{RABIT}  (the look-ahead parameter set to 12) that implements probably the most advanced automata reduction techniques based on computing simulation relations, state merging, and transition pruning \cite{simulation_advanced}.
On these automata, we applied the procedure-finding reduction Algorithm~\ref{alg:main} (with the depth limit $d = 10$), followed by the post-processing optimizations from Section~\ref{sec:post_processing}.

\vspace*{-0.25em}
\paragraph{The Metrics.}

To evaluate the quality of the reduction, we compare the number of states and transitions in the reduced SRA with those in the original NFA, along with the number of introduced register symbols.

We consider both states and register symbols, as they are closely related -- states can be encoded as register symbols and vice versa\footnote{For instance, any NFA can be converted into an equivalent single state SRA with the same number of transitions where the information about the original states is stored in the register and SRA accepts with register equal to some final state of the NFA.}. We, therefore, compare the sum of the number of states and register symbols in the reduced SRA with the number of states in the original NFA.

To determine the reduction in transitions, we compare the number of transitions in the reduced SRA with those in the original NFA. It is important to note that this comparison is not entirely precise because NFA's transitions are of the form $\transition{q}{a}{s}$, while SRA's transitions take the form $\transition{q}{a,\alpha/\beta}{s}$. Nevertheless, we use this metric in the absence of a more precise alternative. While not perfect, it provides a reasonable approximation with a small deviation from the actual reduction.

\vspace*{-0.25em}
\paragraph{Extra Cost of Transitions with Registers.}
For a more concrete analysis, consider the following encoding: a state from $Q$ is stored using 32 bits, a symbol from $\Sigma$ (ASCII character) uses 8 bits, and a register symbol from $\Gamma_{\hspace{-0.5ex}\bot,*}$ requires 4 bits (based on benchmark needs). Using this encoding, an NFA transition would take 72 bits, whereas an SRA transition would take 80 bits (only 11.1\% increase). Moreover, if only SRA transitions that use register symbols were encoded at 80 bits while others remained at 72 bits, the increase would be even smaller (less than 5.6\% in our benchmarks, where more than half of the reduced automaton transitions do not use the register). Therefore, we conclude that comparing the number of transitions provides a reasonable estimate of the actual reduction in automaton size.

\vspace*{-0.25em}
\paragraph{Results Overview.}
The best reduction achieved in the first benchmark was 55.5\% in the number of states and 60.3\% in the number of transitions.
This could significantly improve the performance of the FPGA based methods,
their speed and the maximum number of simultaneously used rules they can handle.
We note that, of course, in order for this result to have a real practical impact, the FPGA based encoding of NFA from \cite{FPGA_based_network_scaning,ApproxRed} has to be generalised to SRA. This is a non-trivial research task beyond the scope of this paper that we plan to address in the future. Nevertheless, we believe that already the current experiment demonstrates that our reduction method works on examples with a~plausible real world motivation.
We achieved similar results on the second benchmark---an average reduction of 51.0\% in states and 34.7\% in transitions. The best reductions recorded were 68.7\% in states and 73.8\% in transitions.

\subsection{Reduction on Snort's Regular Expressions}
\vspace{-3em}

\begin{table}[!htp]
    \caption{Results of applying RABIT/Reduce ($Q_{\mathit{RAB}}, \delta_{\mathit{RAB}}$) and procedure-based reduction ($Q_{\mathit{Map}}+\Gamma'_{\mathit{Map}}, \delta_{\mathit{Map}}$) to seven sets of Snort rules.
    Here, $\Gamma_{\mathit{Map}}$ is the register alphabet obtained from Algorithm~\ref{alg:main} and $\Gamma'_{\mathit{Map}}$ is the final register alphabet after the post-processing.
    The percentages refer to the ratio between the final result and the size of the output of RABIT/Reduce.}
    \label{tab:snort_union}
    \vspace*{0.5em}
    \centering
    \scriptsize
    \setlength{\tabcolsep}{2.5pt}
    \renewcommand{\arraystretch}{1.1}
    \begin{tabular}{|c||r|r||r|r||r|cr|rr|}
        \hline
        Snort rules & $Q_{\mathit{in}}$ & $\delta_{\mathit{in}}$ & $Q_{\mathit{RAB}}$ & $\delta_{\mathit{RAB}}$ & $\Gamma_{\mathit{Map}}$ & \multicolumn{2}{c|}{$Q_{\mathit{Map}} + \Gamma'_{\mathit{Map}}$} & \multicolumn{2}{c|}{$\delta_{\mathit{Map}}$} \\
        \hline\hline
        p2p              & 33  & 1,090  & 32  & 1,084  & 6  &\hspace*{0.5em}$25+2$ &  (-15.6\%) & 570    & (-47.4\%)\\
        worm             & 50  & 3,880  & 34  & 290    & 8  &\hspace*{0.5em}$24+2$ &  (-23.5\%) & 284    &  (-2.1\%)\\
        shellcode        & 162 & 3,328  & 56  & 579    & 2  &\hspace*{0.5em}$48+2$ & (-10.7\%) & 486    & (-16.1\%)\\
        \rowcolor{yellow}
        mysql            & 235 & 30,052 & 91  & 14,430 & 18 & \hspace*{0.5em}$45+5$ & (-45.1\%) & 7,142  & (-50.5\%)\\
        chat             & 408 & 23,937 & 113 & 1,367  & 25 & \hspace*{0.5em}$71+3$ & (-34.5\%) & 1,058  & (-22.6\%)\\
        \rowcolor{yellow}
        specific-threats & 459 & 57,292 & 236 & 31,935 & 32 & \hspace*{0.5em}$99+6$ & (-55.5\%) & 12,680 & (-60.3\%)\\
        telnet           & 829 & 7,070  & 309 & 2,898  & 82 &               $155+4$ & (-48.5\%) & 2,164  & (-25.3\%)\\
        \hline
    \end{tabular}
    \normalsize
\end{table}
\vspace{-1em}

We used regular expressions from seven families of Snort rules and constructed automata from them using the Netbench tool \cite{netbench}. We tested the method on seven automata that were also used in \cite{ApproxRed} created by taking the union of regular expressions from the families. The automata size ranged from 33 to 829 states and from 1,090 to 57,292 transitions. The reduction results for each set of rules, with the two most significant reductions highlighted, are shown in Table \ref{tab:snort_union}.
The best reduction was achieved on the automaton created from the \texttt{specific-threats} rule set. The RABIT/Reduce tool reduced the automaton by 48.6\% in the number of states and by 44.3\% in the number of transitions. The subsequent application of our procedure-finding reduction decreased the number of states by another 55.5\% and the number of transitions~by~60.3\%.

\vspace{-0.5em}
\subsection{Reduction on ARMC, Z3-Noodler, and Regular Expressions}
\vspace{-0.25em}
In this experiment, we tested our method on four benchmarks:
\vspace{0.25em}

\begin{wrapfigure}{r}{0.52\textwidth}
    \vspace*{-2.375em}
    \makeatletter\def\@captype{table}\makeatother
    \caption{The sizes of the benchmark, along with the average and maximum number of states and transitions in the automata.}
    \scriptsize
    \centering
    \setlength{\tabcolsep}{2.5pt}
    \renewcommand{\arraystretch}{1.1}
    \vspace*{-1em}
    \begin{tabular}{|c||r||r|r||r|r|}
                \hline
                Benchmark & \multicolumn{1}{c||}{\#} & $avg(Q)$ & $\mathit{avg}(\delta)$ & $\mathit{max}(Q)$ & $\mathit{max}(\delta)$ \\
                \hline\hline
                ARMC       & 2,604 & 863 & 2,999 & 2,591 &  12,971 \\
                Z3-Noodler &   819 & 63  &   991 &   553 & 104,546 \\
                Email RE   &   362 & 115 & 1,962 &   451 &  24,456 \\
                Param. RE  & 3,656 & 207 & 2,584 &   504 &  12,144 \\
                \hline
            \end{tabular}
            \vspace{-1em}
        \label{tab:benchmarks}
        \normalsize
\end{wrapfigure}
~~
\vspace*{-2em}
\begin{itemize}
    \item The \textit{ARMC} benchmark includes automata generated during Abstract Regular Model Checking \cite{ARMC}.
    \item \textit{Z3-Noodler} includes automata obtained from the Z3-Noodler string solver \cite{z3-noodler-git} during solving the constraints for the SMT Competition\footnote{SMT-LIB: QF\_S/20230329-automatark-lu}.
    \item \textit{Email RE} contains the top 75 regular expressions for email validation from the RegexLib library \cite{RegexLib}, which were combined using the method described in \cite{email1} and \cite{email2} to create 362 automata.
\end{itemize}

\vspace{-1.75em}

\begin{itemize}
    \item The \textit{Param RE} benchmark consists of automata derived from four parametric regular expressions sourced from \cite{TACAS13}.
\end{itemize}

\vspace{-3em}

\begin{table}[!htp]
    \caption{Average and best reduction results in the number of states (Best $Q$) and transitions (Best $\delta$) after applying procedure-based reduction ($Q_{\mathit{Map}} + \Gamma_{\mathit{Map}}$, $\delta_{\mathit{Map}}$) on the results of RABIT/Reduce ($Q_{\mathit{RAB}}$, $\delta_{\mathit{RAB}}$). Here, $\Gamma_{\mathit{Map}}$ is the register alphabet obtained from Algorithm~\ref{alg:main} and $\Gamma'_{\mathit{Map}}$ is the final register alphabet after the post-processing. Percentages indicate the reduction relative to the output of the RABIT/Reduce results.}
    \vspace*{0.5em}
    \centering
    \scriptsize
    \setlength{\tabcolsep}{2.5pt}
    \renewcommand{\arraystretch}{1.1}
    \begin{tabular}{|c|c||r|r||r|r||r|cr|rr|}
        \hline
        $\times$ & Benchmark & $Q_{\mathit{in}}$ & $\delta_{\mathit{in}}$ & $Q_{\mathit{RAB}}$ & $\delta_{\mathit{RAB}}$ & $\Gamma_{\mathit{Map}}$ & \multicolumn{2}{c|}{$Q_{\mathit{Map}} + \Gamma'_{\mathit{Map}}$} & \multicolumn{2}{c|}{$\delta_{\mathit{Map}}$}\\
        \hline\hline

        & ARMC         & 863 & 2,999 & 386 & 1,207 & 124 &              $164+9$     & (-55.2\%) & 1,021 & (-15.5\%)\\
        & Z3-Noodler   &  63 &   991 &  54 &   456 & 11  & \hspace{0.5em}$29+3$     & (-40.7\%) &   276 & (-39.5\%)\\
        & Email RE     & 115 & 1,962 &  93 & 1,759 & 18  & \hspace{0.5em}$48+5$     & (-43.0\%) & 1,256 & (-28.6\%)\\
        \rowcolor{yellow}\cellcolor{white}\multirow{-4}{*}{Average}
        & Param. RE    & 207 & 2,584 &  93 & 1,332 & 6 &  \hspace*{0.5em}$42+3$     & (-51.1\%) &   694 & (-47.9\%)\\
        \hline

        \rowcolor{yellow}\cellcolor{white}\multirow{4}{*}{Best $Q$}
        & ARMC         & 1,518 &  6,397 & 620 & 2,186 & 129 &             $188+6$     & (-68.7\%) & 1,744 & (-20.2\%)\\
        & Z3-Noodler   &   148 & 30,652 & 148 & 2,332 & 9 & \hspace*{0.5em}$43+5$   & (-67.6\%) &   763 & (-67.3\%)\\
        & Email RE\tablefootnote{RABIT/Reduce tool can increase the number of transitions as it tends to make automaton with only one final state.}
        &   223 &  3,553 & 209 & 3,956 & 16 &               \hspace*{0.5em}$70+8$   & (-62.7\%) & 2,643 & (-33.2\%)\\
        & Param. RE    &   359 &    594 & 182 &   417 & 7 & \hspace*{0.5em}$61+4$   & (-64.3\%) &   187 & (-55.2\%)\\
        \hline

        & ARMC         & 459 & 4,445 & 176 &   843 & 62 &   \hspace*{0.5em}$50+6$   & (-68.2\%) &   377  & (-55.3\%)\\
        \rowcolor{yellow}\cellcolor{white}\multirow{-0.5}{*}{Best $\delta$}
        & Z3-Noodler   & 106 &   573 &  95 &   562 & 31 &   \hspace*{0.5em}$50+7$   & (-40.0\%) &   147  & (-73.8\%)\\
        & Email RE     & 263 & 5,793 & 250 & 5,149 & 40 &                 $107+8$   & (-54.0\%) & 2,608  & (-49.3\%)\\
        & Param. RE    & 182 & 7,550 &  63 & 3,716 & 9  &   \hspace*{0.5em}$20+4$    & (-61.9\%) & 1,534  & (-58.7\%)\\
        \hline
    \end{tabular}
    \vspace*{-2em}
    \label{tab:results}
    \normalsize
\end{table}

\begin{figure}[!h]
    \centering
    \captionsetup{justification=justified}
    \captionsetup[subfigure]{justification=centering}
    \begin{subfigure}[h]{0.24\textwidth}
        \centering
        \includegraphics[width=1.025\textwidth, trim=17 17 17 0, clip]{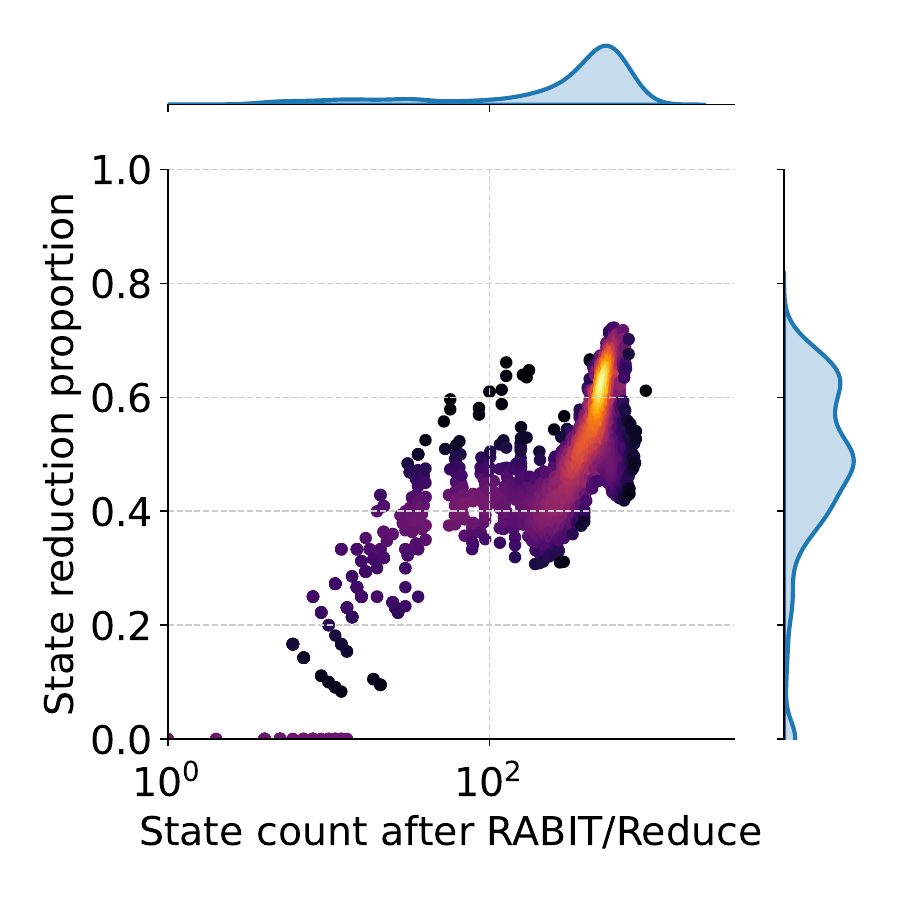}
        \vspace*{-1.5em}
        \caption*{ARMC - $Q$}
    \end{subfigure}
    \begin{subfigure}[h]{0.24\textwidth}
        \centering
        \includegraphics[width=1.025\textwidth, trim=17 17 17 0, clip]{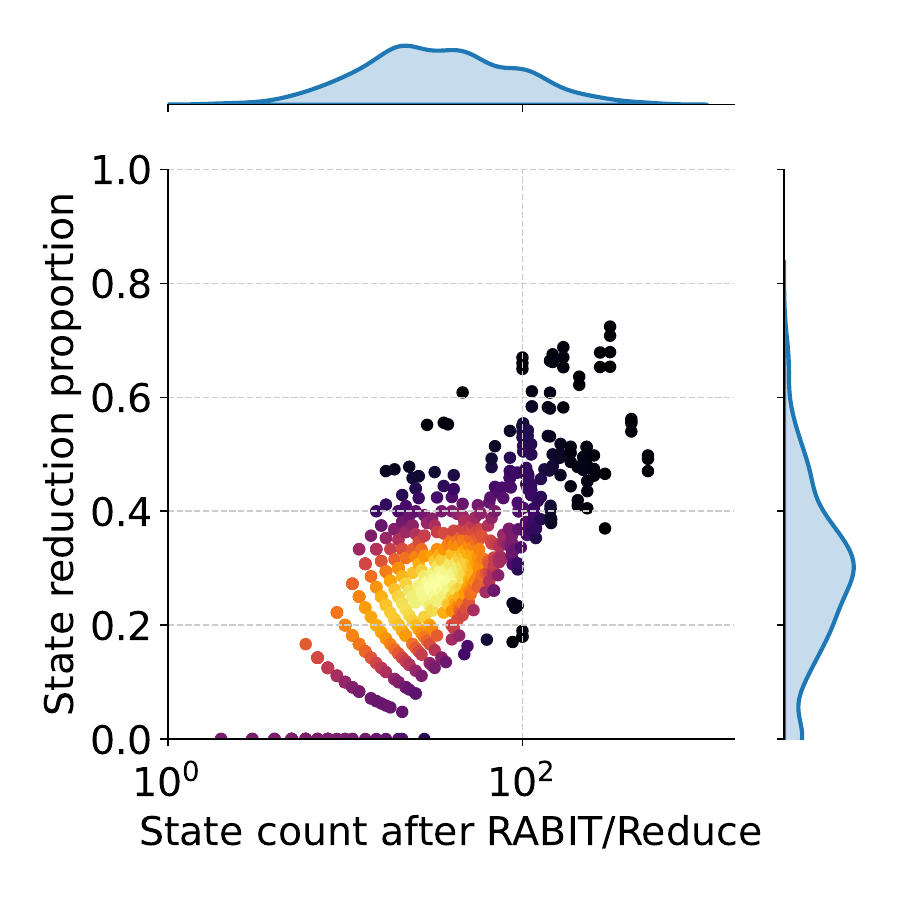}
        \vspace*{-1.5em}
        \caption*{Z3-Noodler - $Q$}
    \end{subfigure}
    \begin{subfigure}[h]{0.24\textwidth}
        \centering
        \includegraphics[width=1.025\textwidth, trim=17 17 17 0, clip]{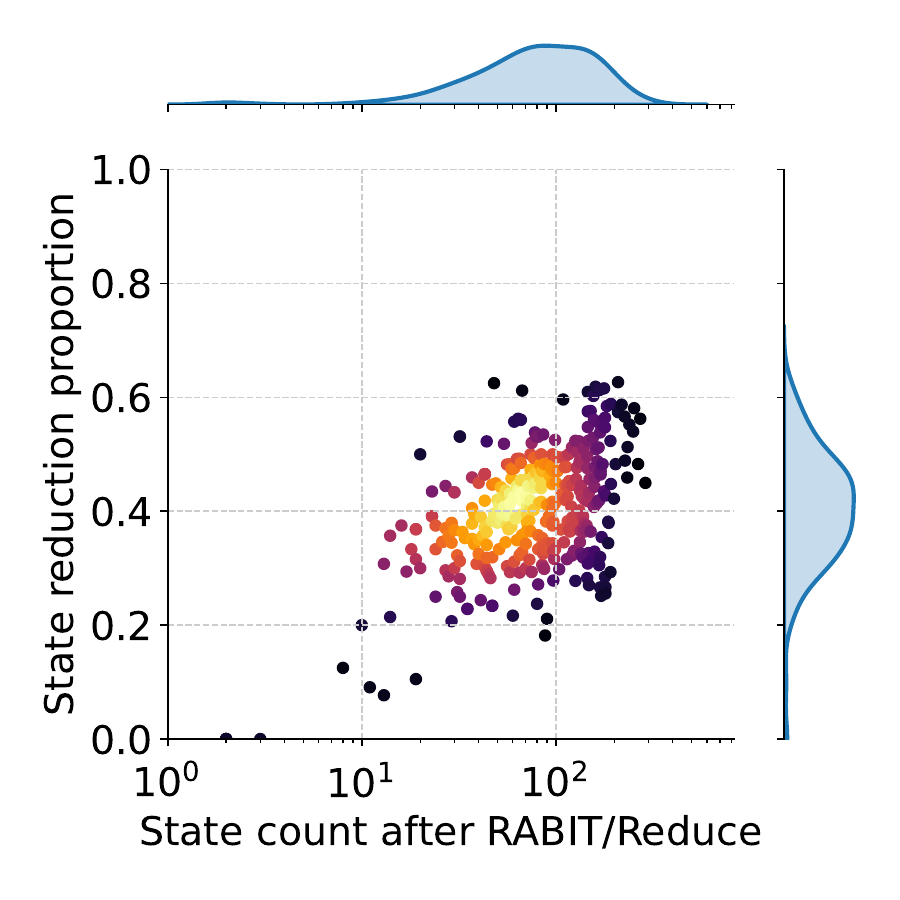}
        \vspace*{-1.5em}
        \caption*{Email RE - $Q$}
    \end{subfigure}
    \begin{subfigure}[h]{0.24\textwidth}
        \centering
        \includegraphics[width=1.025\textwidth, trim=17 17 17 0, clip]{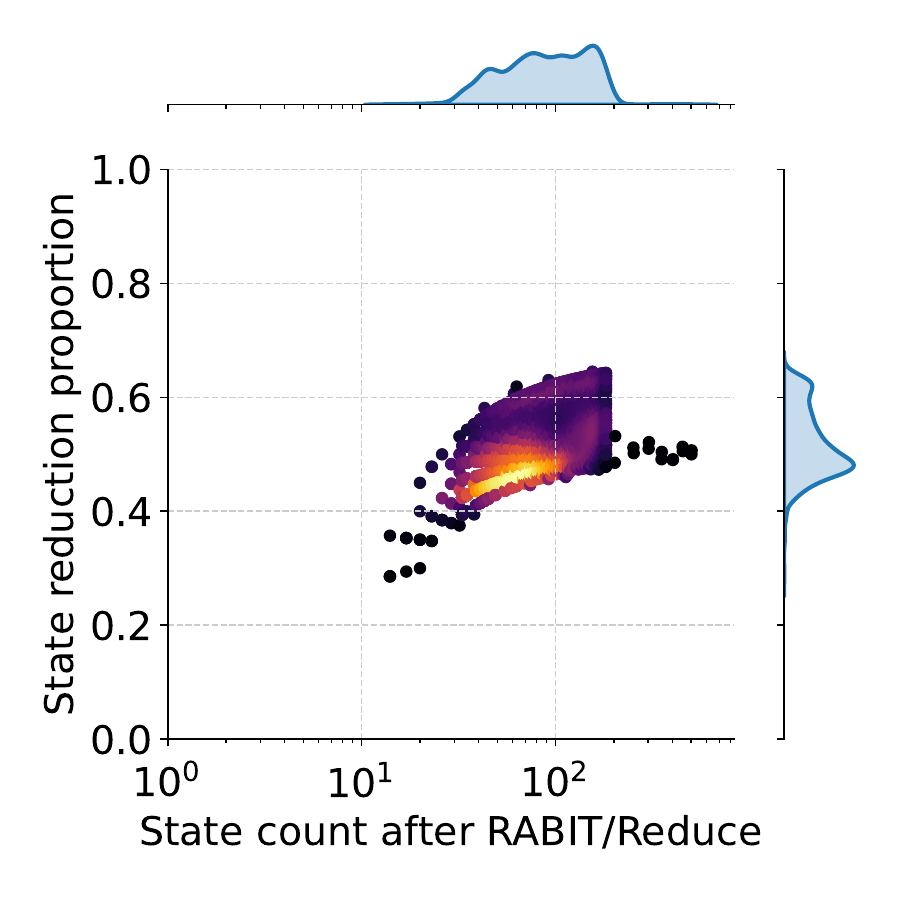}
        \vspace*{-1.5em}
        \caption*{Param. RE - $Q$}
    \end{subfigure}

    \begin{subfigure}[h]{0.24\textwidth}
        \centering
        \includegraphics[width=1.025\textwidth, trim=17 17 17 0, clip]{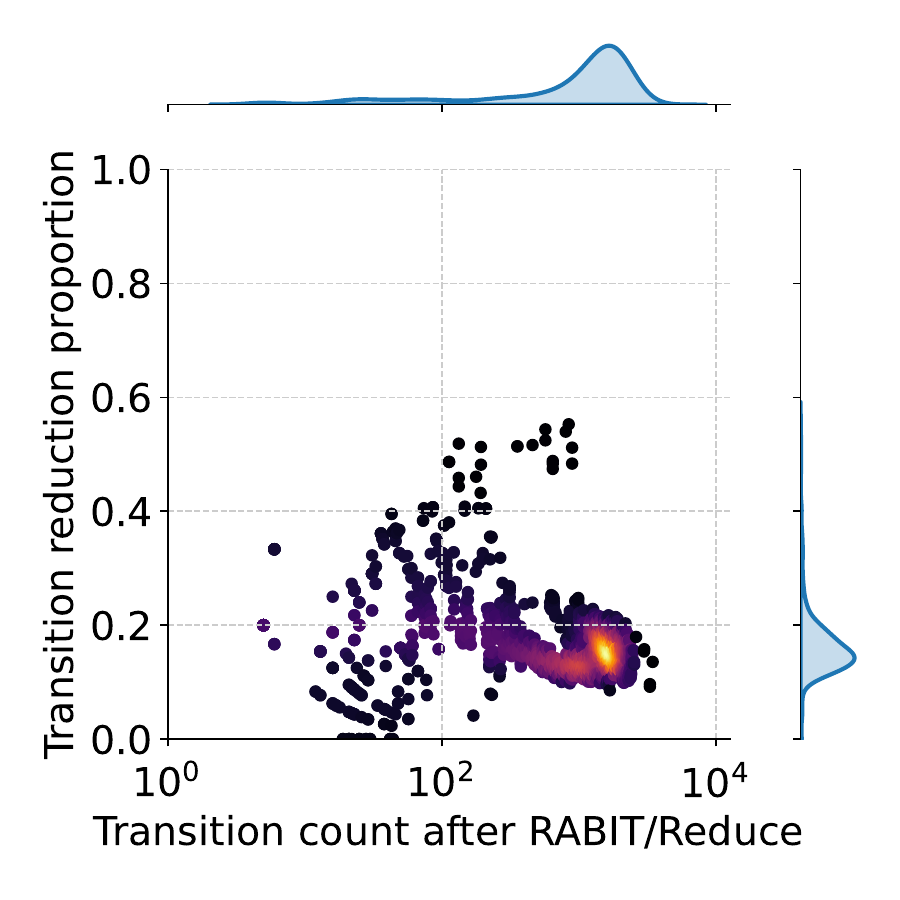}
        \vspace*{-1.5em}
        \caption*{ARMC - $\delta$}
    \end{subfigure}
    \begin{subfigure}[h]{0.24\textwidth}
        \centering
        \includegraphics[width=1.025\textwidth, trim=17 17 17 0, clip]{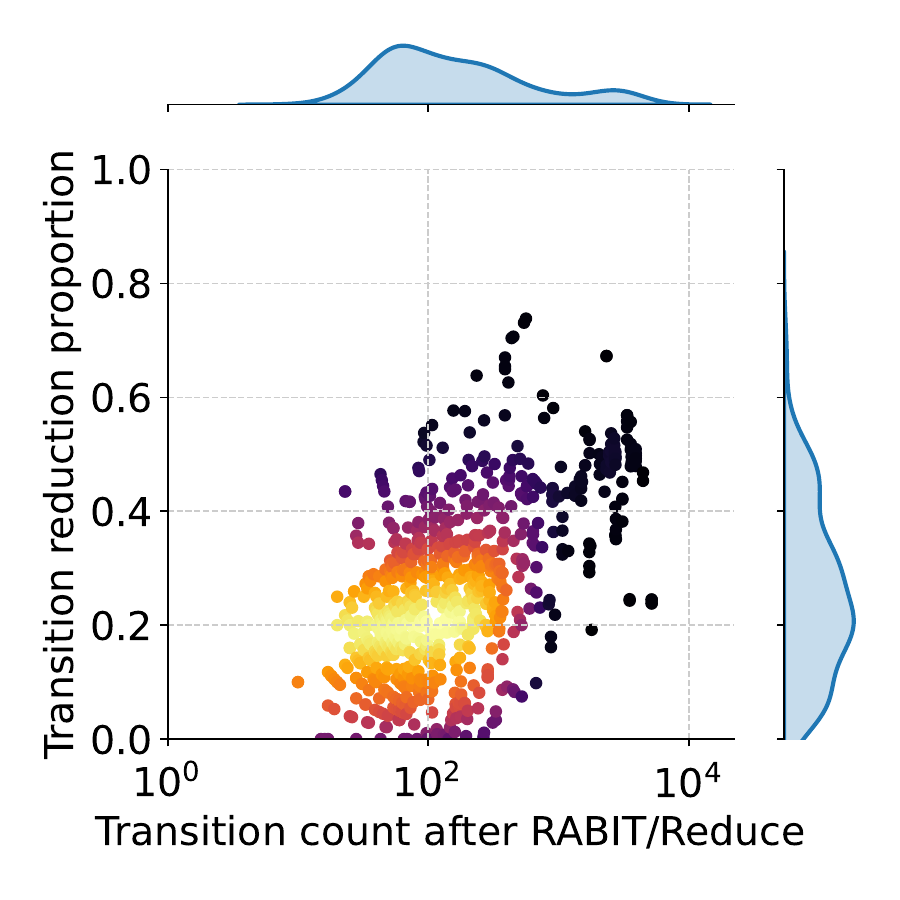}
        \vspace*{-1.5em}
        \caption*{Z3-Noodler - $\delta$}
    \end{subfigure}
    \begin{subfigure}[h]{0.24\textwidth}
        \centering
        \includegraphics[width=1.025\textwidth, trim=17 17 17 0, clip]{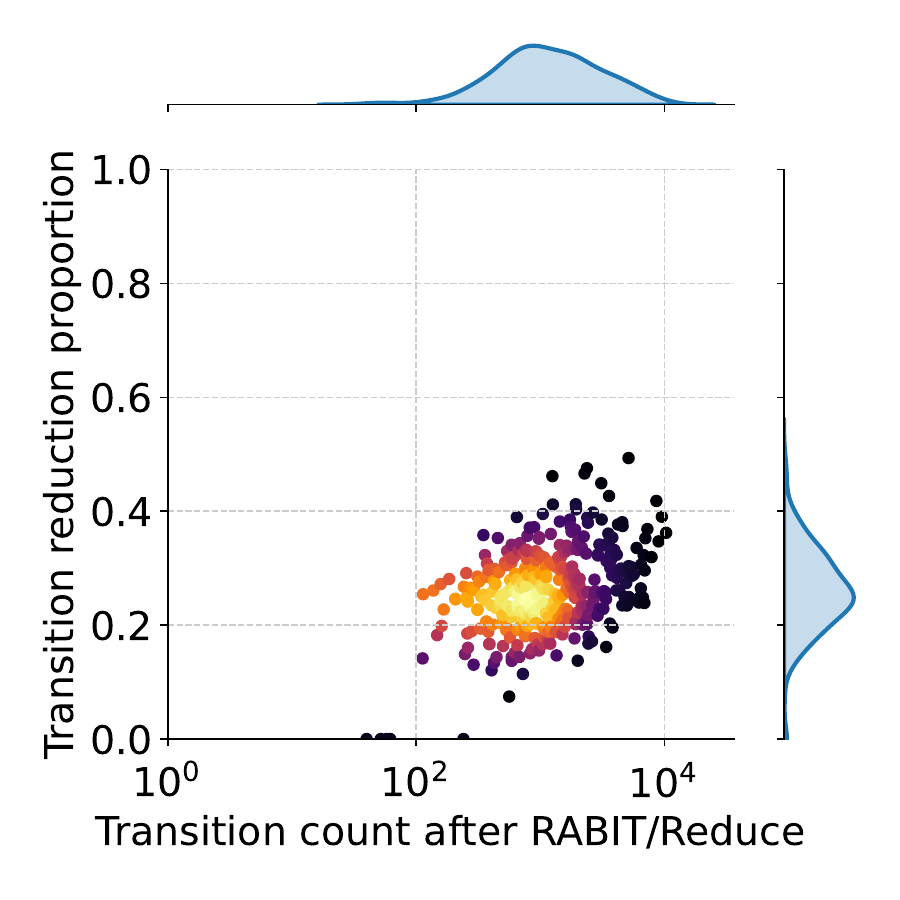}
        \vspace*{-1.5em}
        \caption*{Email RE - $\delta$}
    \end{subfigure}
    \begin{subfigure}[h]{0.24\textwidth}
        \centering
        \includegraphics[width=1.025\textwidth, trim=17 17 17 0, clip]{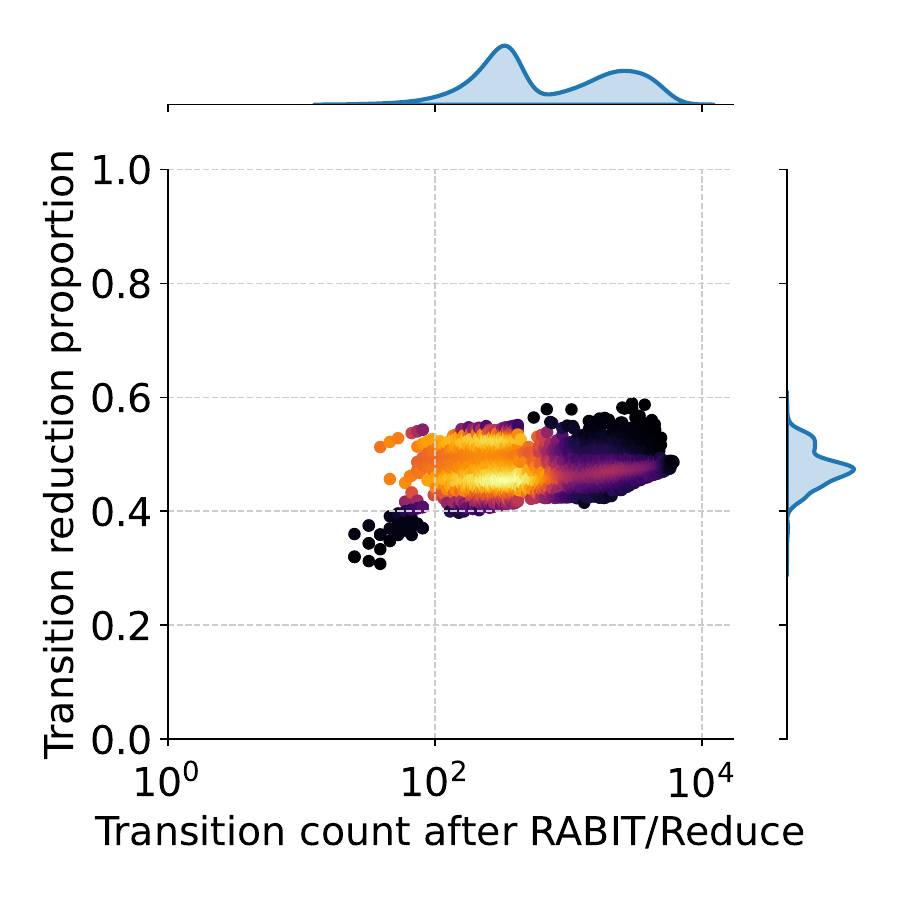}
        \vspace*{-1.5em}
        \caption*{Param. RE - $\delta$}
    \end{subfigure}
    \vspace*{-0.5em}

    \caption{The relation between the achieved size reduction and the original size, in terms of states ($Q$) and transitions ($\delta$), for ARMC, Z3-Noodler, Email RE, and Parametric RE benchmarks (on automata reduced by RABIT/Reduce).}
    \label{fig:results}
\end{figure}

The average and best reduction results for each benchmark are shown in Table~\ref{tab:results}.
Figure \ref{fig:results} visualises the dependency of the achieved reduction on the size of automata.
On average (thought of all automata sets), the number of states was reduced by 51\% and the number of transitions by 34.7\%.
The greatest reduction in the number of states was achieved with the automaton from the ARMC benchmark, which showed a reduction of 68.7\% in states and 20.2\% in transitions. Conversely, the most significant reduction in transitions was observed in the automaton from the Z3-Noodler benchmark, achieving a reduction of 40\% in states and 73.8\% in transitions.

The results on this benchmark show that our reduction method has good results across various domains and sources of automata.

\vspace*{-0.75em}

\vspace{-2.5em}
\section{Conclusions}
\vspace{-0.25em}
We have introduced a novel automata size reduction technique based on representing multiple similar sub-graphs as a single procedure that can be called from multiple contexts, and uses a finite domain register to remember the context.
We have provided technical details of an implementation of this idea.
Our implementation proved to be effective on automata from various applications,
especially on automata used in hardware accelerated regular pattern matching,
where our size reduction may potentially significantly improve the efficiency.

This work represents a first take on this idea, its main point being that it can be very effective and practically relevant.
It can be further elaborated in many directions, e.g.,
generalizing the shape of procedures that we can infer beyond linear;
researching more efficient algorithms of finding the procedures;
exploring and comparing various other types of memory that could be used to store call context, especially stacks and bit-vectors.
Another topic is providing alternatives to the basic algorithms for NFAs that will work with automata with procedures symbolically, and implementing these in automata libraries.
We also plan to extend the presented experiment with the automata from the FPGA-accelerated pattern matching by providing an efficient FPGA implementation of our automata with register-implemented procedures.

\bibliographystyle{splncs04}
\bibliography{paper}

\end{document}